\documentclass[]{hdsr}

\graphicspath{{figs/}}

\begin{document}

\newgeometry{bottom=1.5in}


\begin{center}
\LARGE{\textbf{Time-varying exposure history and subsequent health outcomes: a two-stage approach to identify critical windows}}
  \title{} 
  \maketitle

  \thispagestyle{empty}

  \begin{tabular}{cc}
  \normalsize{
    Maude Wagner\upstairs{\affilone}, 
    Francine Grodstein\upstairs{\affiltwo}, 
    Karen Leffondre\upstairs{\affilone},} 
    \normalsize{C\'ecilia Samieri\upstairs{\affilone,*},
    C\'ecile Proust-Lima\upstairs{\affilone,*}} 
    \\ [0.25ex]
   \small{\upstairs{\affilone} Inserm, Bordeaux Population Health Research Center, UMR U1219,} \normalsize{Univ. Bordeaux, France} \\
   \small{\upstairs{\affiltwo} Rush Alzheimers Disease Center, Rush University Medical Center, Chicago, IL, USA} \\
   \small{\upstairs{*} These authors contributed equally} 
  \end{tabular}
  
  \emails{~~~~~~~~~~~~      Corresponding author: maud.wagner@u-bordeaux.fr}
  \vspace*{0.1in}

\begin{abstract}
\footnotesize{
Long-term behavioral and health risk factors constitute a primary focus of research on the etiology of chronic diseases. Yet, identifying critical time-windows during which risk factors have the strongest impact on disease risk is challenging. To assess the trajectory of association of an exposure history with an outcome, the weighted cumulative exposure index (WCIE) has been proposed, with weights reflecting the relative importance of exposures at different times. However, WCIE is restricted to a complete observed error-free exposure whereas exposures are often measured with intermittent missingness and error. Moreover, it rarely explores exposure history that is very distant from the outcome as usually sought in life-course epidemiology. We extend the WCIE methodology to (i) exposures that are intermittently measured with error, and (ii) contexts where the exposure time-window precedes the outcome time-window using a landmark approach. First, the individual exposure history up to the landmark time is estimated using a mixed model that handles missing data and error in exposure measurement, and the predicted complete error-free exposure history is derived. Then the WCIE methodology is applied to assess the trajectory of association between the predicted exposure history and the health outcome collected after the landmark time. In our context, the health outcome is a longitudinal marker analyzed using a mixed model. A simulation study first demonstrates the correct inference obtained with this approach. Then, applied to the Nurses' Health Study (19,415 women) to investigate the association between body mass index history (collected from midlife) and subsequent cognitive decline (evaluated after age 70), the method identified two major critical windows of association: long before the first cognitive evaluation (roughly 24 to 12 years), higher levels of BMI were associated with poorer cognition. In contrast, adjusted for the whole history, higher levels of BMI became associated with better cognition in the last years prior to the first cognitive interview, thus reflecting reverse causation (changes in exposure due to underlying disease). In conclusion, this approach, easy to implement, provides a flexible tool for studying complex dynamic relationships and identifying critical time windows while accounting for exposure measurement errors.}
\end{abstract}
\end{center}

  \small	

\section{INTRODUCTION}
\begin{figure}[ht]
\centering
\includegraphics[width=12cm]{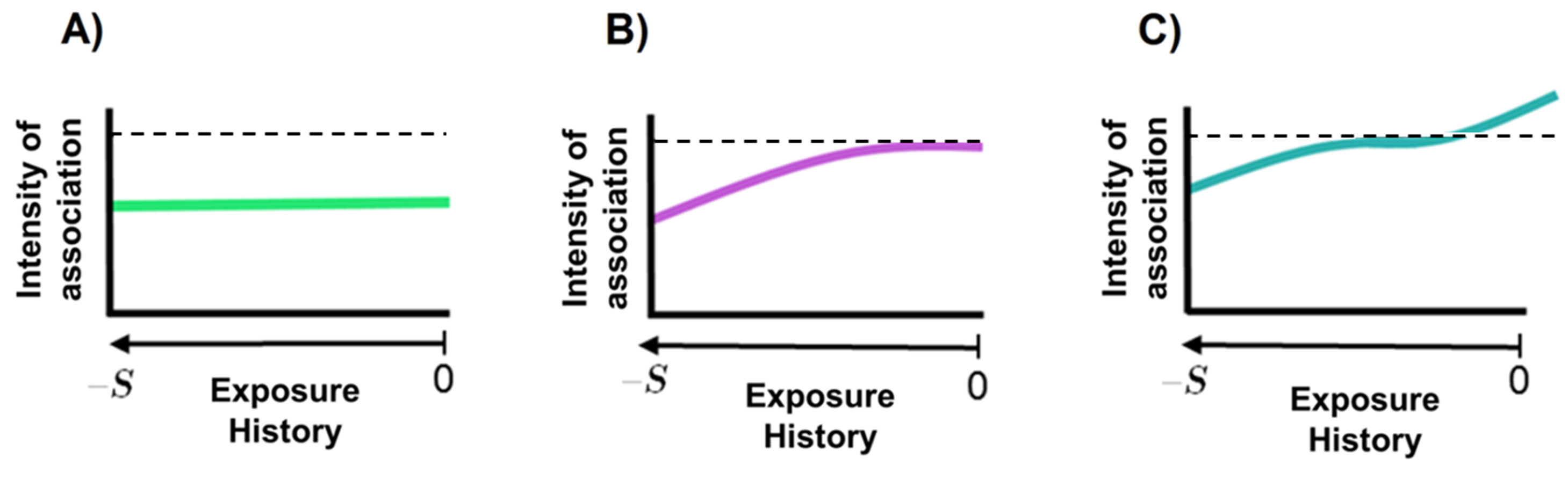}
\caption{\small Three examples of trajectories of association between an exposure history from time $-S$ to time 0 and a subsequent health outcome: (A) constant association, (B) time-varying with remote association or (C) time-varying with opposing remote and recent associations. The dashed horizontal lines represent the 0 association.} 
\label{Figure1}
\end{figure}

\noindent
Long-term lifestyle, environmental or occupational exposures may have a major impact on the occurrence of chronic diseases. Yet, identifying critical time-windows during which risk factors have the strongest impact on disease risk is challenging. Chronic diseases often result from a long and accumulating pathological process that evolves over years before diagnosis (Liu et al., 2010). In such context, the exposure time-windows close to the clinical event may be less meaningful in terms of etiology and, importantly, may be obscured by reverse causality (which occurs when behaviors or exposures change as the disease progresses in infra-clinic stages). For example, while obesity in midlife is a risk factor for subsequent dementia, it is consistently reported as a protective factor later in life; indeed advanced neuropathology, by altering olfactory function and taste, potentially leads to malnutrition and weight loss in late-life (Tolppanen et al., 2014; Singh-Manoux et al., 2018; Wagner et al., 2018; Wagner et al., 2019). Thus, the only valuable approach to evaluate causal associations linking cumulative unhealthy body weight to cognitive aging might be to estimate the relationship between the entire history of exposure that precedes and begins well upstream of the period at risk of the event. 
\\
Exposure history metrics have been developed to estimate the cumulative effect of time-varying exposure on disease endpoints (Checkoway et al., 1992; Stranges et al., 2006). However, to be relevant in life-course epidemiology, such methodologies absolutely need to handle: (i) associations that can vary according to the distance between the exposure and the disease endpoint, (ii) exposures that are measured only at sparse time points and with error, and (iii) exposure and disease endpoint time-windows that do not necessarily coincide. Finally, they should also apply to longitudinal disease outcomes (not only survival or binary) and their change over time such as cognitive decline that is a strong predictor of dementia. This work specifically aimed at extending the methodology of exposure history metrics to address all these important issues.  
\\
The most common exposure history metric is the cumulative index of exposure (CIE) (Checkoway et al., 1992; Stranges et al., 2006). Computed as the un-weighted sum of past exposures, CIE assumes that past values of exposure are all of equivalent importance (see Figure \ref{Figure1}, Scenario A) which may induce etiologically inadequate conclusions when the effect of the exposure on health outcomes likely varies according to the age or timing of exposures (see Figure \ref{Figure1}, Scenarios B and C). To address this challenge, Breslow et al. (Breslow et al., 1983) and Thomas (Thomas et al.,1988) introduced the concept of weighted CIE (WCIE), that combines information about duration, intensity and timing of the exposure. WCIE represents the time-weighted sum of past exposures, with weights reflecting the relative importance of exposures at different times. Therefore, unlike CIE, the estimated effects of the exposure history on health status are time-varying, a key aspect for the identification of critical windows of exposure in life-course epidemiology. Weights can be assigned a priori by a known parametric weight function based on clinically relevant assumptions (Vacek et al., 1997; Langholz et al., 1999; Abrahamowicz et al., 2006) or they can be directly estimated from the data using a flexible spline-based regression (Eilers et al., 2015). WCIE models are found in a wide range of applications with binary and time-to-event outcomes, such as occupational asbestos in relation to mesothelioma (Lacourt et al., 2017, Leveque et al., 2018) or drug use linked to fall-related injuries (Sylvestre et al., 2012).

However, both CIE and WCIE methodologies always require the exposure history to be complete. Yet, intermittent missing data on individual exposures are frequently reported in case-control and cohort studies, in particular when follow-up is long, thus preventing any analysis using these metrics. Second, although measurement error is inevitable, CIE and WCIE metrics also require the exposures to be measured without error and the violation of this assumption (which is likely in most epidemiological contexts) may lead to biased exposure-risk relationships (Prentice et al., 1982). To overcome these limitations, Mauff et al. (Mauff et al., 2017) introduced a weighted cumulative association structure within a joint model to assess the impact of an endogenous time-varying exposure measured intermittently and with error on the risk of a time-to-event outcome. However, in this approach, exposure and outcome have to be evaluated during the same time-window, while as introduced previously, in some contexts the interest is in the cumulative effect of an exposure history that precedes evaluation of disease risk. Finally, all these methodologies have been developed for survival or binary outcomes. To date, we are aware of only one pharmaco-epidemiological study applying the WCIE in association with a longitudinal outcome (Danieli et al., 2020). Nonetheless, it considered a complete error-free exposure history measured concomitantly with the outcome, and it only assessed its association with the level of the longitudinal outcome while in etiological studies, the interest lies in the change over time of the longitudinal outcome.  
\\
In this manuscript, we propose to extend the WCIE methodology to address life-course epidemiology research questions. We rely on a landmark approach which consists in limiting the analysis of the outcome of interest after a certain time of interest, called landmark, and considering as predictors the exposure history up to the landmark. Although mainly developed in dynamic prediction context (Ferrer et al., 2019, Houwelingen et al., 2007), it also addresses the issue of non-concurrency between the exposure and the outcome time-windows central in life-long epidemiology. 
\\
In Section 2, we first introduce the proposed methodology, and describe the estimation procedure. In Section 3, we apply the method in the Nurses' Health Study to investigate the association between BMI history starting in midlife and repeated cognition assessed after age 70. In Section 4, we demonstrate that the estimation procedure provides correct inference in a simulation study. Finally we discuss the results and methodology, and conclude.

\section{Methods}

\subsection{A landmark approach for assessing the dynamics of association between an intermittently evaluated time-varying exposure and a subsequent health outcome}

\subsubsection{General study framework}

\begin{figure}[ht]
\centering
\includegraphics[width=12cm]{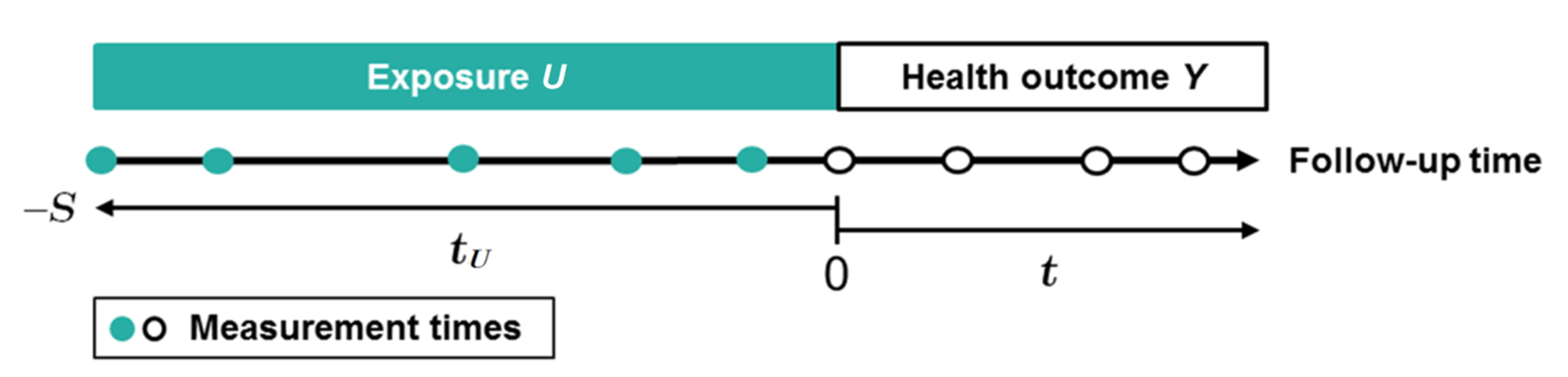}
\caption{\small Temporal representation of the non-concomitant measurement times of the exposure and the subsequent health outcome considered in our study framework. For both exposure and health outcome, measurements are collected at discrete and individual-specific occasions with error. The exposure history is constructed according to the time $t_U$ prior to the first outcome assessment at time 0 ($-S \le t_U \le 0$). The longitudinal health outcome $Y$ is modeled prospectively according to time $t~(t \ge 0)$.} 
\label{Figure2}
\end{figure}

Within a prospective cohort study, we consider a sample of N individuals present in the cohort at a landmark time of interest from which the health outcome is to be studied, hereafter called time 0 (see Figure \ref{Figure2}). The temporal framework of the study consists in repeated measures of the exposure prior to the landmark time, that is from a negative time $-S$ (e.g. time of entry in the cohort) up to time 0, as well as repeated measures of the outcome collected at and after the landmark time. Following the landmark framework (Ferrer et al., 2019), exposures data collected after time 0, if any, are not included. Both exposure and health outcome are collected at discrete and individual-specific occasions with error. The design of the Nurses’ Health Study used in the application (Section 3) naturally follows the landmark framework since the exposure was assessed several decades before the first assessment of outcome. In this manuscript, we will focus mainly on a longitudinal heath outcome. Nevertheless, the proposed approach also applies to other types of event, such as time-to-event and binary endpoints.

\subsubsection{Mixed model for the exposure}

Let $U_{il}$ denote the exposure values collected for individual $i$ ($i$ = 1, ..., $N$) at the retrospective measurement time $t_{Uil}~(l = 1, ..., m_i)$ before the landmark time so that $t_{Uil} \in [-S,0]$ (see Figure \ref{Figure2}). The times and the total number $m_i$ of repeated measures can differ from one individual to the other thus implicitly handling intermittent missing values. Although the methodology could apply beyond, we consider here a continuous exposure and describe its trajectory via a linear mixed effects model (Laird et al., 1982): 

\begin{equation}
\label{eq_expo}
\begin{split}
~~~~~~~~~~~~~~~~~~~~U_{il} &= U^{*}_i(t_{Uil}) + \varepsilon_{il}
\\
&= X_i(t_{Uil})^\top\beta + Z_i(t_{Uil})^\top b_i + \varepsilon_{il}
\end{split}
\end{equation}

where $U^{*}_i(t_{Uil})$ is the true underlying exposure value at measurement time $t_{Uil}$, $X_i(t_{Uil})$ and $Z_i(t_{Uil})$ are two vectors of covariates at time $t_{Uil}$ associated with the vectors of fixed effects $\beta$ and of random effects $b_i$, respectively; $b_i \sim \mathcal{N}(0,B)$ and is independent of $\varepsilon_{il}$, the Gaussian measurement error with mean 0 and variance $\sigma^{2}_\varepsilon$. The objective of this linear mixed model is to accurately model the underlying true exposure $U^{*}_i(t_U)$ at any time $t_U$ in $[-S,0]$ for inclusion in the WCIE. As such, both $X_i(t_U)$ and $Z_i(t_U)$ should include flexible functions of time, for instance a basis of splines (Eilers et al., 2015) or fractional polynomials (Royston et al., 1999). 

\subsubsection{Weighted cumulative index of exposure}

WCIE is defined as the weighted sum of the underlying true exposures during the period of interest. Without loss of generality, we consider the entire window $[-S,0]$ although any sub-window could be considered instead: 

\begin{equation}
\label{WCIE}
~~~~~~~~~~~~~~~~~~~~~~~~WCIE_{i} = \sum \limits_{t_U=-S}^0 w(t_U)U^{*}_i(t_U)
\end{equation}

where $w(t_U)$ is the weight function assigned to the history of the true exposure $U_i^{*}(t_U)$ during the $S+1$ time units (e.g., years) preceding the initial assessment of the outcome (see Figure  \ref{Figure1}). 

\subsubsection{Model for the health outcome} 

Although transposable to cross-sectional or time-to-event outcomes, our primary interest was to estimate the time-varying effects of a past exposure history on a subsequent longitudinal health outcome (see Figure \ref{Figure2}). Thus let’s denote $Y_{ij}$ the measure for individual $i$ $(i = 1, ..., N)$ collected at time $t_{ij} \ge 0$ with $j = 1, ..., n_i$. The times and the total number $n_i$ of repeated measures can differ from one individual to the other. Change over time of $Y$ is modeled using a linear mixed model (Laird et al., 1982). For ease of presentation, we consider here a linear trajectory over time, and thus introduce two time-varying effects of the exposure, one on the level at the landmark time (denoted $WCIE_I$) and one on the change over time of $Y$ (denoted $WCIE_S$):  

\begin{equation}
\label{eq_MLM}
\begin{split}
Y_{ij} =~&\alpha_0~+~\tilde{X_i}(t_{ij})^\top \alpha_1~+~ WCIE_{Ii}~\gamma_I~+~c_{0i}~+~\\
 \biggl(&\alpha_2~+~\tilde{X_i}(t_{ij})^\top \alpha_3~+~ WCIE_{Si}~\gamma_S~+~c_{1i} \biggr)\times t_{ij}~+~\tilde{\varepsilon}_{ij}\\
 \newline 
=~& \alpha_0~+~\tilde{X_i}(t_{ij})^\top \alpha_1~+~ \sum \limits_{t_U=-S}^0 \gamma_I~w_I(t_U)U^{*}_i(t_U)~+~c_{0i}~+~\\
 \biggl(&\alpha_2~+~\tilde{X_i}(t_{ij})^\top \alpha_3~+~ \sum \limits_{t_U=-S}^0 \gamma_S~w_S(t_U)U^{*}_i(t_U)~+~c_{1i} \biggr)\times t_{ij}~+~\tilde{\varepsilon}_{ij}\\
\end{split}
\end{equation}

where $\tilde{X_i}(t_{ij})$ is a vector of covariates at time $t_{ij}$ associated with the vectors of fixed effects $\alpha_1$ and $\alpha_3$; $WCIE_{Ii}$ and $WCIE_{Si}$ are the weighted cumulative exposure covariates associated to the initial level and to the slope of $Y$, respectively. They are defined according to Equation \eqref{WCIE}  with different weights $w_{I}(t_U)$ and $w_{S}(t_U)$, and are associated with fixed effects $\gamma_I$ and $\gamma_S$; $c_{0i}$  and $c_{1i}$ are correlated individual random intercept and slope, respectively, with $c_i \sim \mathcal{N}(0,\tilde{B})$; $\tilde{\varepsilon}_{ij}$ are the independent Gaussian measurement errors with mean zero and variance $\sigma^{2}_{\tilde{\varepsilon}}$.

\subsubsection{Identification of the trajectories of association}

The linear mixed model of the health outcome defined in Equation \eqref{eq_MLM} provides two trajectories of association (i.e., time-varying effects) of the history of past exposure on the subsequent health outcome, one for the initial level (noted $\gamma_I^{*}(t_U)$), one for the slope (noted $\gamma_S^{*}(t_U)$):

\begin{equation}
\label{eq_gamma}
\begin{split}
~~~~~~~~~~~~~~~~~~~~~~~~~~~~~~\gamma_I^{*}(t_U) &= \gamma_I w_I(t_U)
\\
\gamma_S^{*}(t_U) &= \gamma_S w_S(t_U)
\end{split}
\end{equation}
\\
They represent the mean difference of initial level (i.e. $\gamma^{*}_{I}(t_U)$) or of change per unit of time (i.e. $\gamma^{*}_{S}(t_U)$) when the exposure increases of 1-unit at time $t_U$, adjusted for other covariates and when the exposure history is stricly similar at any other time in $[-S,0]$.

\subsubsection{Specification of weights}

In some applications, the form of the weight function $w_.(t_U)$ (where subscript . refers either to $I$ or $S$ in Equations \eqref{eq_MLM} and \eqref{eq_gamma}) is known. However, most often, it has to be estimated directly from the data. As others previously (Danieli et al., 2020, Hauptmann et al., 2000, Sylvestre et al., 2009), we chose to estimate the weight function by regression using a basis of splines (Eilers et al., 2015), which are flexible enough to capture a variety of clinically plausible shapes (Smith et al., 1979). Thus, the weight function can be written:
\begin{equation}
\label{eq_3}
~~~~~~~~~~~~~~~~w_.(t_U) = \theta_{.0} + \sum \limits_{k=1}^K \theta_{.k} B_{k}(t_U) = \sum \limits_{k=0}^K \theta_{.k} B_{k}(t_U) 
\end{equation}

where $(B_{k})_{k=1,...K}$ refers to the $K$ basis of splines functions and $(\theta_{.k})_{k=1,...K}$ the coefficients to estimate. For ease of calculation, we denote the intercept $\theta_{.0} B_{0}(t_U)$ with $B_{0}(t_U)=1$, $\forall~t_U$.

Although any type of splines could be considered, we favored natural cubic splines that limit erratic behavior at the boundary of the time window thanks to linearity constraints  (Wood et al., 2017). In that case, the basis of splines is built from $K+1$ knots which are to be chosen inside the time window $[-S, 0]$. The number of knots has to be carefully determined from the data. A large number of knots implies high flexibility but it may lead to overfitting. On the contrary, a small number of knots may result in an oversmooth estimate that is prone to under-fit bias (Wood et al., 2017). In our work, we considered between one and five inner knots (i.e., $K \in [2,6]$) and relied on the Akaike information criterion (Akaike et al., 1978) to select the final splines basis. In addition, in the absence of prior knowledge, we considered equidistant knots.
\\
One additional advantage of approximating the weight function using splines is that the WCIE can be rewritten as a linear combination of $K+1$ components:

\begin{equation}
\label{WCIEspl}
\begin{split} 
~~~~~~~~~~~~~~~~WCIE_{.i} &= \sum \limits_{t_U=-S}^0 w.(t_U)U^{*}_i(t_U)\\ 
&=\sum \limits_{t_U=-S}^0 \sum \limits_{k=0}^K \theta_{.k} B_{k}(t_U) U^{*}_i(t_U) \\
&= \sum \limits_{k=0}^K \theta_{.k} \underbrace{\sum \limits_{t_U=-S}^0 B_{k}(t_U) U^{*}_i(t_U)} \\
&= \sum \limits_{k=0}^K \theta_{.k} \hspace*{1.2cm} H_{ki}
\end{split}
\end{equation}

where $H_{ki}$ (for $k$=0,…,$K$) denote intermediate summary covariates of the exposure history (Abrahamowicz et al., 1996, Sylvestre et al., 2009, Danieli et al., 2020).

\subsubsection{Identifiability constraints}

The time-varying effects of the exposure history defined in Equation \eqref{eq_gamma} is overparameterized; coefficients  $\gamma_I$ and  $\gamma_S$, and the weights $w_I(t_U)$ and $w_S(t_U)$ cannot be simultaneously estimated from the data without further constraint. Hauptmann et al. (Hauptmann et al., 2000) proposed in a case-control study to estimate the total effect of the history of exposures $\gamma_.$ and constrain $w_.(t_U)$ with the constraint $\sum \limits_{t_U=-S}^0 w_.(t_U) = S+1$. Following Sylvestre et al. (Sylvestre et al., 2009) and Danieli et al. (Danieli et al., 2020), we directly estimated the time-varying coefficients without constraining the weights by setting the coefficients $\gamma_I$ and $\gamma_S$ to 1. Therefore, the time-varying effects directly correspond to the weights and Equation \eqref{eq_gamma} becomes:

\begin{equation}
\label{eq_gamma2}
\begin{split}
~~~~~~~~~~~~~~~~~~~~~~~~~~~~~~~~~\gamma_I^{*}(t_U) &= w_I(t_U) \\
\gamma_S^{*}(t_U) &= w_S(t_U)
\end{split}
\end{equation}
\\
With this parameterization, the overall mean effect of the history of exposure is $\frac{1}{S+1}\sum \limits_{t_U=-S}^0 \gamma_I^{*}(t_U) = \frac{1}{S+1}\sum \limits_{t_U=-S}^0 w_I(t_U)$ and $\frac{1}{S+1}\sum \limits_{t_U=-S}^0 \gamma_S^{*}(t_U) = \frac{1}{S+1}\sum \limits_{t_U=-S}^0 w_S(t_U)$.

Finally, by considering an approximation of the weight functions by splines (Equation \eqref{WCIEspl}) and by not constraining the weights (Equation \eqref{eq_gamma2}), the linear mixed model defined in Equation \eqref{eq_MLM} becomes:

\begin{equation}
\label{eq_MLMspl}
\begin{split}
~~~~~~Y_{ij} =~&\alpha_0~+~\tilde{X_i}(t_{ij})^\top \alpha_1~+~\sum \limits_{k=0}^K \theta_{Ik}H_{ki}~+~ c_{0i}~+~\\
 \biggl(&\alpha_2~+~\tilde{X_i}(t_{ij})^\top \alpha_3~+~\sum \limits_{k=0}^K \theta_{Sk}H_{ki}~+~ c_{1i} \biggr)\times t_{ij}~+~\tilde{\varepsilon}_{ij}
\end{split}
\end{equation}
\\
where $\theta_{Ik}$ and $\theta_{Sk}$ are $K+1$ unconstrained parameters to be estimated, and $H_{ki}$ are the intermediate covariates defined from \eqref{WCIEspl}.

\subsection{Maximum Likelihood Estimation using a two-stage procedure}

By sharing the underlying true exposure level ${U}^{*}_i(t_U)$, the sub-model for the exposure in Equation \eqref{eq_expo} and the health outcome model in Equation \eqref{eq_MLM}, define a shared parameter joint model, which would call for the simultaneous estimation of all parameters in order to avoid any bias (Tsiatis et al., 2004). However, given the peculiar temporal framework of the landmark approach with the inclusion of subjects present in the cohort at the landmark time and the distinct windows of observation for the two variables, a two-stage estimation procedure is unlikely to generate any bias – this assumption is confirmed by simulations in Section 4. 
\\
Let $\phi_U = (\beta, \sigma_\varepsilon, \text{vec}(B))$ and $\phi_Y = (\alpha, \theta_{.1}, ..., \theta_{.K}, \sigma_{\tilde{\varepsilon}}, \text{vec}(\tilde{B})$) denote the vectors of parameters to be estimated in the linear mixed model for the exposure (in Equation \eqref{eq_expo}) and in the linear mixed model for the outcome (in Equation \eqref{eq_MLMspl}), respectively.

\subsubsection{First stage}
In the first stage, we classically compute the maximum likelihood estimators of $\phi_U$. Then, based on the estimated fixed effects parameters  $\widehat{\beta}$ and the best linear unbiased predictors of the random effects $\widehat{b_i}$, the individual-specific true exposure can be predicted at any time $t_U$ preceding the initial assessment of the outcome:
\begin{equation}
\label{eq_preE}
~~~~~~~~~~~\widehat{U}^{*}_i(t_U) = X_i(t_{U})^\top\widehat{\beta} + Z_i(t_{U})^\top \widehat{b_i}~~~~~~~~~~~~\forall~t_U \in [-S,0]
\end{equation}

\subsubsection{Second stage} 

In the second stage, the true underlying exposure level $U^{*}_i(t_U)$ is replaced by its individual estimation $\widehat{U}^{*}_i(t_U)$. Therefore, in the case of an approximation by splines (as introduced in Section 2.1.6.), $WCIE_{.i} = \sum \limits_{k=0}^K \theta_{.k} H_{ki}$ in Equation \eqref{WCIEspl} is replaced by $\widehat{WCIE}_{.i} = \sum \limits_{k=0}^K \theta_{.k} \widehat{H}_{ki}$ with $\widehat{H}_{ki} =\sum \limits_{t_U=-S}^0 B_{k}(t_U) \widehat{U}^{*}_i(t_U)$. The model for the outcome thus becomes:

\begin{equation}
\label{eq_MLMspl2}
\begin{split}
~~~~~~Y_{ij} =~&\alpha_0~+~\tilde{X_i}(t_{ij})^\top \alpha_1~+~\sum \limits_{k=0}^K \theta_{Ik} \widehat{H}_{ki}~+~c_{0i}~+~\\
 \biggl(&\alpha_2~+~\tilde{X_i}(t_{ij})^\top \alpha_3~+~\sum \limits_{k=0}^K \theta_{Sk} \widehat{H}_{ki}~+~ c_{1i} \biggr)\times t_{ij}~+~\tilde{\varepsilon}_{ij}
\end{split}
\end{equation}
\\
and parameters $\phi_Y$ can be again estimated by classical likelihood maximization. 
\\
In the end, the estimated time-varying effects of the pre-landmark-time exposure on the level of the outcome at the landmark time (i.e. $\widehat{\gamma}^{*}_I$) and on the change over time of the outcome after the landmark time (i.e. $\widehat{\gamma}^{*}_S$) are:
\begin{equation}
\begin{split}
\label{eq_gamm_est}
~~~~~~~~~~~~~~~~~~~~~~~~\widehat{\gamma}{^{*}}_{I}(t_U) &= \widehat{w_I}(t_U) = \sum \limits_{k=0}^K \widehat{\theta}_{Ik} B_{k}(t_U)\\
\widehat{\gamma}{^{*}}_{S}(t_U) &= \widehat{w_S}(t_U) = \sum \limits_{k=0}^K \widehat{\theta}_{Sk} B_{k}(t_U)
\end{split}
\end{equation}

\subsubsection{Standard error estimation}

One drawback of two-stage estimation approach is that the variances of the parameters obtained in the second stage do not account for the variability of the parameters estimated in the first stage. To properly take into account this variability of the first stage, we used parametric bootstrap (Efron et al., 1993): instead of including in the second stage $\widehat{U^*}(t_U)$ computed at the maximum likelihood estimate $\widehat{\phi_U}$, we generated $M$ bootstrap replicates $\phi_{Um}$ (for $m=1, ...,M$) from the asymptotic distribution $\mathcal{N}(\widehat{\phi_U},\widehat{V(\widehat{\phi_U}}))$ and computed the corresponding $\widehat{U^*}_m(t_U)$ to be included in the second stage where $\widehat{\phi_{Ym}}$ and $\widehat{V(\widehat{\phi_{Ym}})}$ were then obtained. The total variance $\widehat{V_{tot}(\widehat{\phi_Y})}$ of the  estimated parameters $\widehat{\phi_Y}$ was obtained as the sum of the intra- and inter-replicate variances:

\begin{equation}
\label{var_tot}
\widehat{V_{tot}(\widehat{\phi_Y})} = \frac{1}{M} \sum \limits_{m=1}^M  \widehat{V(\widehat{\phi_{Ym}})} + \frac{1}{M} \sum \limits_{m=1}^M (\widehat{\phi_{Ym}}-\overline{\widehat{\phi_{Ym}}})(\widehat{\phi_{Ym}}-\overline{\widehat{\phi_{Ym}}})^\top
\end{equation}

The variance of $\widehat{\gamma.}(t_U)$ was then easily derived using the Delta-method (Boos et al., 2013).

\subsubsection{Implementation}

The methodology can be implemented in any standard statistical software. We chose \textbf{R} programming (version 3.6.0) and function \textbf{hlme} of the \textbf{R} package \textbf{lcmm} (version 1.7.8) (Proust-Lima  et al., 2017) for estimating the linear mixed models. The codes of the simulation studies are openly available in GitHub at https://github.com/MaudeWagner/WHistory.
\newline

\section{Application to body mass index history starting at mid-life and subsequent cognitive trajectories after age 70}

To emphasize the utility of our methodology, we investigated in a prospective cohort study the relationship of BMI history collected since mid-life with subsequent cognitive function and cognitive decline in older ages, where prior data indicate changing relations over time, with the possibility of reverse causation at older age (Singh-Manoux et al., 2018, Wagner et al., 2018, Wagner et al., 2019). 

\subsection{The Nurses’ Health Study}

\begin{figure}[ht]
\centering
\includegraphics[width=11cm]{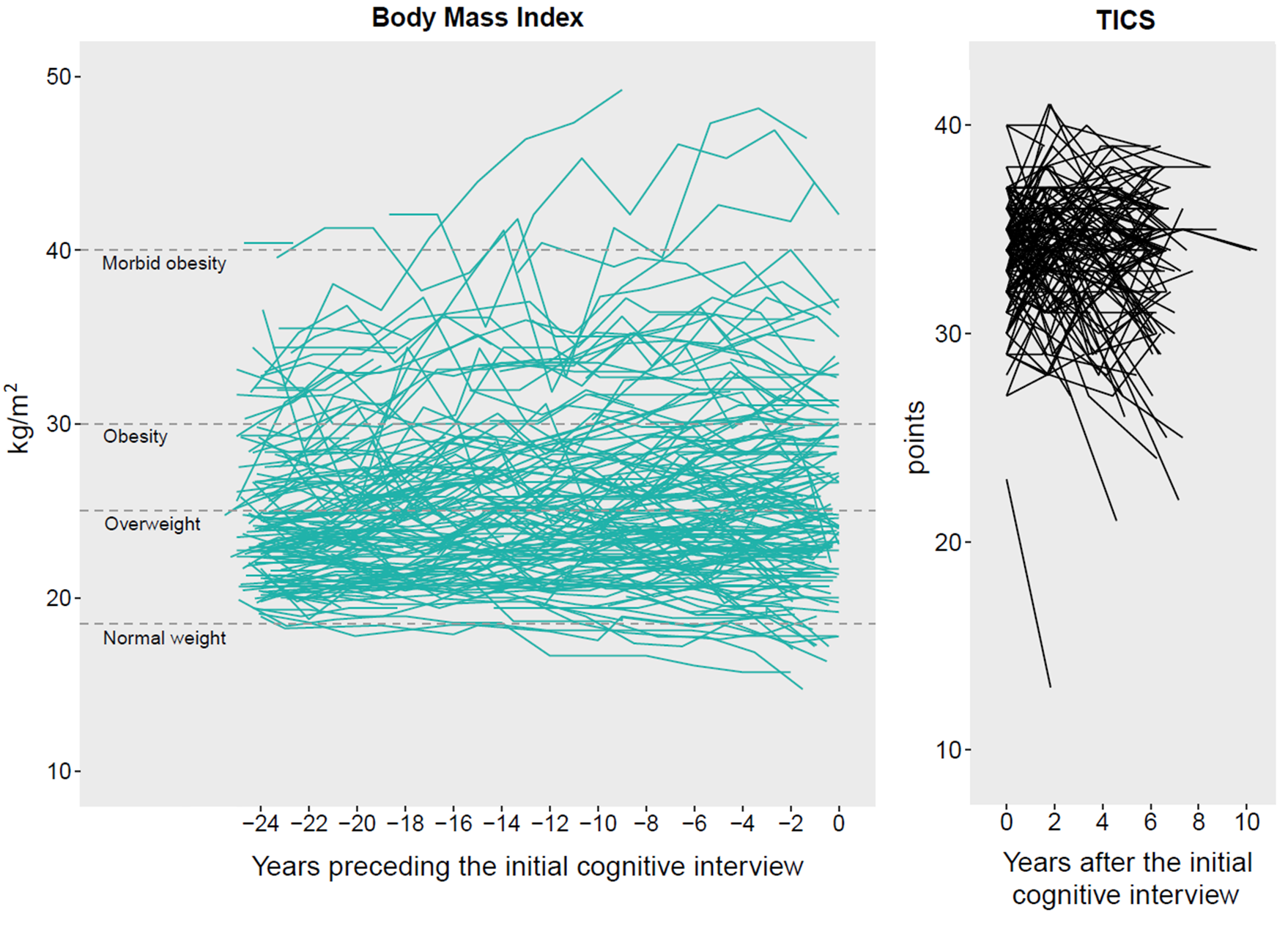}
\caption{\small Observed individual trajectories of body mass index in the 24 years of the window of exposure preceding the first cognitive interview (left panel) and of Telephone Interview for Cognitive Status (TICS) score over the window of cognitive assessment (right panel) for 150 randomly selected women from Nurses' Health Study, United States (1976-2008).}
\label{Figure3}
\end{figure}

We relied on data from the Nurses’ Health study (NHS). NHS began in 1976, when 121,700 female registered nurses, aged 30-55 years and residing in 11 US states, returned a mailed questionnaire about their lifestyle and health, including their weight and height (Colditz et al., 1997). Thereafter, the participants continued to complete biennial questionnaires, with updated data. BMI was computed as self-reported weight in kilograms divided by height in meters squared; self-reported weight was highly correlated with measured weight in a previous validation study in 184 participants (Willett et al., 1983). From 1995 to 2001, all nurses who had reached age 70 or older with no history of stroke were invited to participate in a telephone-based study of cognitive function; the entry in this sub-study corresponds to our landmark time. Among eligible women, 19,415 (92\%) completed the initial validated Telephone Interview for Cognitive Status (TICS, score range 0 to 41), a telephone-based modified version of the Mini-Mental State Examination (Folstein et al., 1975). Cognitive assessments were repeated at three occasions approximately every 2 years, with a high follow-up rate (>90\% among those remaining alive at each follow-up point). 
\\
Following our landmark framework (see Figure \ref{Figure2}), we considered the history of BMI from the entry in the cohort through the first assessment of TICS (at time 0); more than 95\% of participants had information on BMI up to 24 years before the initial cognitive interview. For the current analyses, among the 19,415 participants of the cognitive sub-study, we only excluded 34 women who did not have at least one measure of BMI between enrollment and the first cognitive interview or with missing data for educational level (an important potential confounder), leading to a study sample of 19,381 women. 
\\
At enrollment, the mean age of women was 50.5 years-old (SD=2.5), 77.9\% had the registered nurse diploma as highest educational level and 34.9\% were overweight or obese whereas only 1.3\% were underweight. At the landmark time of the initial cognitive interview, mean age was 73.3 years-old (SD=2.3) a majority of women was overweight or obese (53.4\%), average TICS was 33.7 points (SD=2.8) and 11.2\% of nurses had cognitive impairment. On average, 11.4 (SD=1.7) BMI measurements per person were collected over the 23.7 years (SD=1.2) of the window of exposure, followed by 3.2 (SD=1.1) TICS measurements collected subsequently over 4.9 years (SD=2.5). Figure \ref{Figure3} shows the observed individual trajectories of BMI in the 24 years of the window of exposure preceding the first cognitive interview and those of the TICS in the 8 years of the window of cognitive assessment in a subsample of 150 randomly selected women. In general, BMI tended to increase over time while TICS decreased.

\subsubsection{Specification of the statistical models}

We considered the following linear mixed effect models for the BMI and the subsequent TICS score:

\begin{equation}
\label{lcl}
\begin{array}{ccc}
 \text{BMI}_{il} = &\beta_0 + \beta_1 \text{age0}_i + \beta_2 \text{education} _i + F(t_{Uil})^\top \beta_3 + b_{0i} + F(t_{Uil})^\top b_{1i} + \varepsilon_{il}  
\\
 \\
\text{TICS}_{ij} = &\alpha_0 + \alpha_1\text{age0}_i + \alpha_2 \text{education}_i + \sum \limits_{k=1}^K \theta_{Ik}H_{ki} + c_{0i} + \alpha_3 \text{V0}_{ij}~ + 
 \\
 &\biggl(\alpha_4 + \alpha_5 \text{age0}_i + \alpha_6 \text{education}_i + \sum \limits_{k=1}^K \theta_{Sk}H_{ki} + c_{1i}\biggr)\times t_{ij} + \tilde{\varepsilon}_{ij}
 \end{array}
\end{equation}
where $\text{age0}$ is the age of women at enrollment into the cognitive sub-study in years; $\text{education}$ is the highest degree of diploma (binary variable: Registered nurse versus Bachelor's, Master's or Doctorate); $\text{V0}$ is an indicator for the first cognitive assessment which captures a possible first-passing effect (Vivot2016); $F(t_{Uil})$ is a basis of natural cubic splines with 4 inner knots located at 20th, 40th, 60th and 80th percentiles of the elapsed time of exposure (the number of knots was determined by the AIC);  and $H_{ki}$ are the intermediate covariate summarizing BMI history: $H_{ki} = \sum \limits_{t_U=-S}^0 B_{k}(t_U) \text{BMI}_i^{*}(t_U)$. The weight function of the WCIE was approximated using natural cubic splines with $K-1=2$ inner knots located at 33th and 66th percentiles (the number of knots was determined by AIC). $\beta$, $\alpha$, $b_i$, $c_i$, $\varepsilon_{il}$, and $\varepsilon_{ij}$ have been previously defined in Section 2. In order to facilitate the interpretation of the results and as introduced in the methods, we assumed linear trajectories of cognitive decline. This choice seemed reasonable since cognition was evaluated in the NHS over a short follow-up.

\subsection{Results}

\begin{figure}[ht]
\centering
\includegraphics[width=11cm]{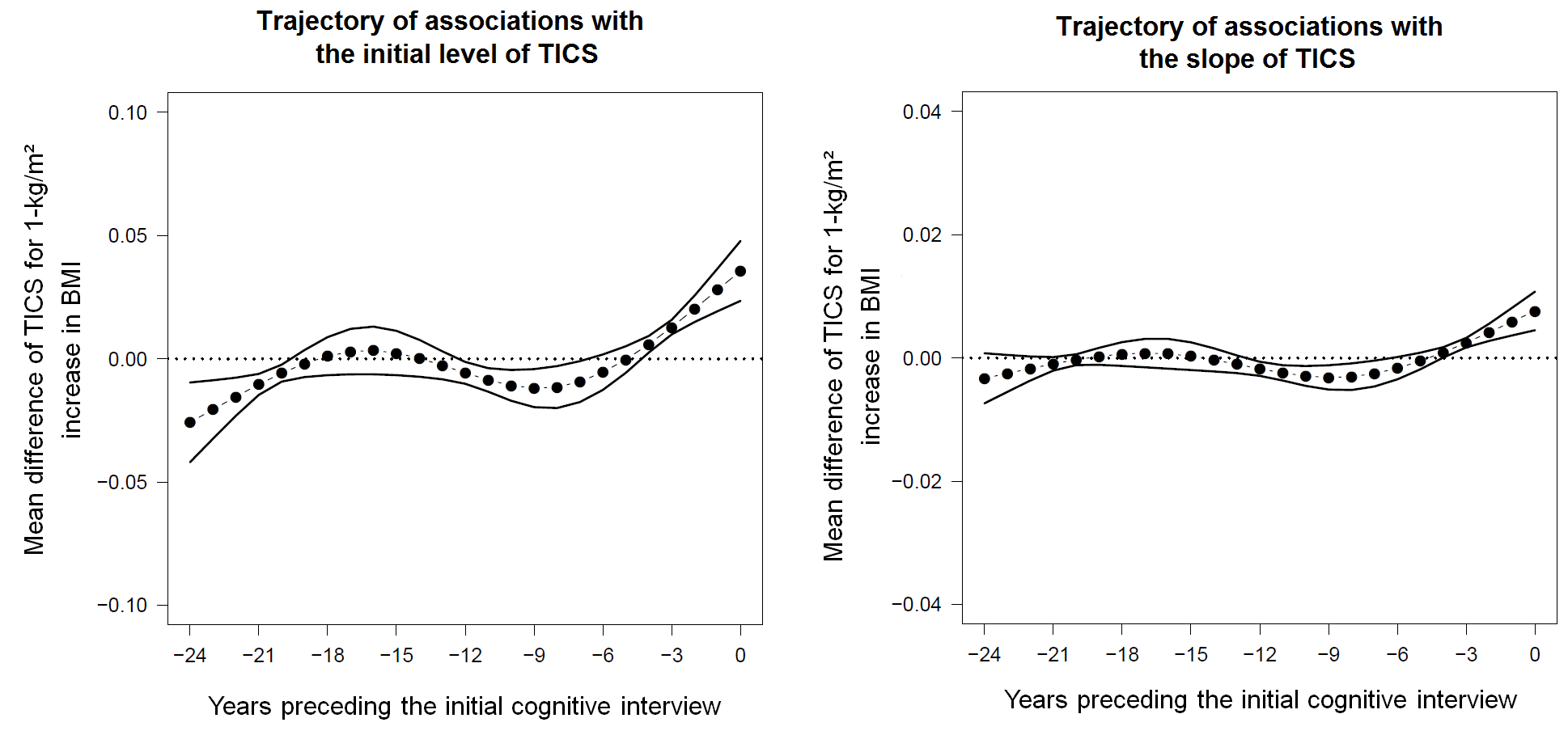}
\caption{\small Trajectories of associations between body mass index (BMI) history in the 24 years prior to the first cognitive interview on the initial level (left panel) or the change with time (right panel) of the Telephone Interview for Cognitive Status (TICS) score in the Nurses' Health Study (N=19,381), United States (1976-2000). 95\% confidence intervals were obtained by parametric bootstrap with 500 replicates. A negative estimate indicates that increased BMI is related to worse cognition/more cognitive decline and a positive estimate indicates better cognition/less cognitive decline.}
\label{Figure4}
\end{figure}

Figure \ref{Figure4} represents the trajectory of association of BMI pre-landmark-time history in the 24 years with the initial level of TICS (left panel) and its change over time (right panel). Each point represents the association coefficient of BMI at a specific year adjusted for age, educational level, and all BMI history at any other year.

The overall mean association of BMI history over the whole 24 year period of exposure was significant with a negative relation to cognition for both the initial level (-0.0013 [95\% CI: -0.0014;-0.0012]) and the annual slope of decline in TICS (-0.00016 [95\% CI:-0.00019;-0.00015]). In addition, as expected, the association between BMI exposure history and cognition was non-constant, with shapes of associations on both the initial level and the slope of TICS suggesting opposite remote and recent effects. For the initial level (see Figure \ref{Figure4}, left panel), higher BMI from -24 to -20 years prior to cognitive assessment, corresponding to mid-life, were significantly associated with lower mean TICS at the first cognitive interview assessed after age 70. For example, for similar confounders (i.e., age and education level) and BMI trajectories at any times except at -24 years, a 1-kg/m2 increase of BMI was associated with a lower initial TICS level of 0.025 point on average (95\% CI = -0.039; -0.009). Between -20 and -12 years, for similar confounders and BMI trajectories at any times except at the year evaluated, BMI levels were no longer significantly associated with initial TICS level whereas, from -12 to -5 years, higher BMI levels were again associated with worse cognitive function, consistently with that observed earlier in midlife. In contrast, starting around 5 years before cognitive assessment, higher BMI levels became significantly associated with higher initial level of TICS (likely reflecting reverse causation). 

For the slope of TICS (see Figure \ref{Figure4}, right panel), results showed no significant association with BMI levels from -24 to -13 years preceding the first cognitive interview. However, higher BMI levels between -12 and -5 years and lower BMI levels between -5 years and the first cognitive interview were associated with worse cognitive decline, similarly to what observed with the initial level of TICS (but to a lesser extent).

\begin{figure}[ht!]
\centering
\includegraphics[width=10cm]{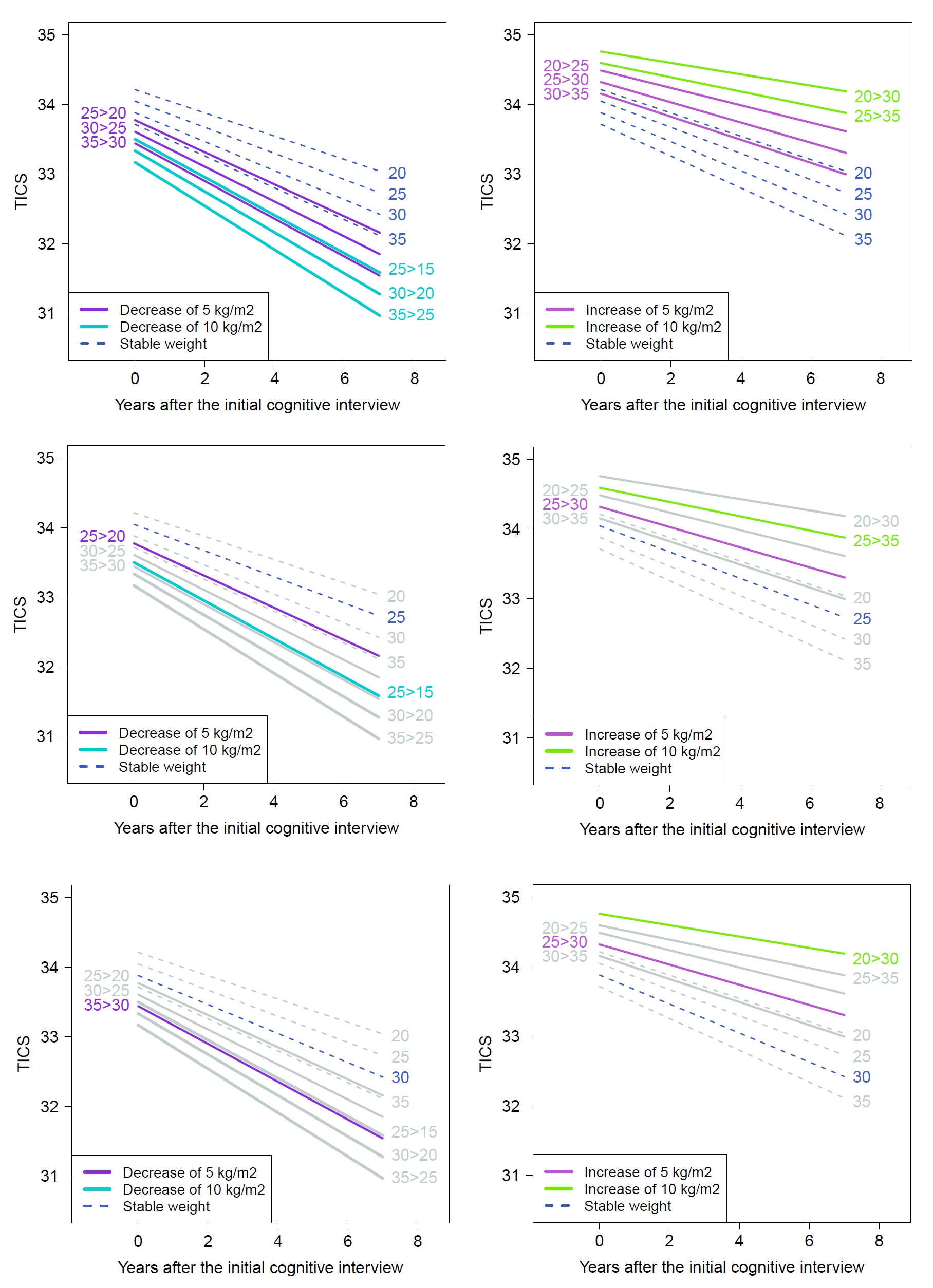}
\caption{\small Mean change of the TICS score over the study course following 5 theoretical BMI history profiles defined all over the whole 24 years preceding the first cognitive interview: (i) linear decrease of 5 points of BMI  (purple lines), (ii) linear decrease of 10 points of BMI (blue lines), (iii) stable BMI (dashed lines), (iv) linear increase of 5 points of BMI (pink lines), (v) linear increase of 10 points of BMI (green lines). For example, on the top left panel, women who had a BMI of 25kg/m2 24 years prior to the first cognitive assessment and linearly dropped to a BMI of 20kg/m2 at the initial cognitive interview (i.e. legend labelled "25>20") have an initial average TICS level of 33.8 points that decreases over time to 32.4 points after 7 years of follow-up.} 
\label{Figure5}
\end{figure}

To help interpret these complex trajectories of association, we examined hypothetical BMI trajectories (e.g., stable BMI of 25 kg/m2 over the 24 years of exposure) in relation to subsequent changes in TICS (see Figure \ref{Figure5}). As observed with the trajectories of association, these predictions by profile of BMI generally showed that women with greater BMI over time had higher (better) initial mean levels of TICS compared to women who had a stable or a decreasing BMI with age, reflecting as expected the major influence of reverse causation. Overall, this application provides additional evidence that relationships between BMI and cognition are complex and largely depend on careful consideration of the window of exposure when BMI is assessed. 
When considering piecewise constant weights instead of weights approximated by natural cubic splines, the trajectories of time-varying effects remained similar (see eFigure 1) suggesting that the splines function well reflected the data. Moreover, results remained generally the same after adjusting the linear mixed model of the outcome for additional potential confounders (i.e., chronic disease history, smoking status or postmenopausal hormone therapy) categorized as binary variables and collected over the same period as BMI history. Likewise, we observed similar trajectories of associations of BMI on cognitive function when we stratified analyses for these factors (results not shown).

\newpage
\section{Simulation study}
\subsection{Overview of the simulation design}
To validate our methodology and its two-stage estimation procedure, we generated cohort data that mimicked the NHS design and temporal framework of Figure \ref{Figure2}, according to a joint model with the shared true exposure as introduced in Section 2.1. In the main simulation study, we generated individual-specific visit times for exposure, using a uniform distribution in [-6, 6] months around theoretical visits every 2 years from -24 years to landmark time 0. At each visit time, we simulated the observed value of the exposure $U$ according to a mixed model as defined in Equation \eqref{lcl} and considered a 10\% proportion of missing visits completely at random (as observed in NHS). The trajectory of $U$ was modelled according to natural cubic splines (with 4 inner knots at 20th, 40th, 60th and 80th percentiles of the observation times) at both the population and individual levels, and was adjusted for two covariates mimicking age at study entry (normal distribution: mean 51, standard deviation 3) and binary education level (0.25-probability Bernoulli distribution).

We also generated individual-specific visit times for the outcome with a uniform distribution in [-6, 6] months around theoretical visits every 2 years from 0 to 7 years. We then generated the values of the health outcome $Y$ according to the mixed model as defined in Equation \eqref{lcl} adjusted for age and education (on both the intercept and the slope) and the indicator for the first outcome assessment, and considering a 3\% proportion of missing visits completely at random (as observed in NHS). In this model, the true underlying exposure level was considered.

\subsection{Scenarios}

Parameters in these generating models corresponded roughly to those obtained in the application data for BMI and TICS (parameters provided in eTables 1 and 2) with exception for the trajectory of association with BMI history for which we assumed three different relevant scenarios plotted in Figure \ref{Figure1}:

\begin{itemize}
    \item Scenario A: constant negative association over the time window of exposure (this corresponds to a standard CIE) with $\gamma_I(t_U)=-0.05$ and $\gamma_S(t_U)=-0.01$ for all $t_U \in [-24,0]$; 
    
    \item Scenario B: negative association far upstream of the outcome assessment and null association at the approach of the initial assessment using a truncated centred normal distribution:
    $\gamma_I(t_U) = (\Phi (\frac{{t_U}+24}{6}) - 1)*0.1$ and $\gamma_S(t_U) = (\Phi (\frac{{t_U}+24}{6}) - 1)*0.03$ with $t_U \in [-24,0]$;
    
    \item Scenario C: trajectories of association obtained in the application between BMI history and TICS trajectory (see Figure \ref{Figure4}).
\end{itemize}

In the main simulations, we considered a framework close to NHS with frequent exposure data (roughly every two years), a small proportion of missing visits (10\% for the exposure and 3\% for the outcome), and a small measurement error for the exposure ($\sigma_\varepsilon$ = 0.9). However, to further evaluate the methodology in less favorable situations, we also considered cases where (i) the exposure was measured every 4 years, (ii) the proportion of missing data was larger for the exposure (20\%), and (iii) the error of measurement was larger ($\sigma_\varepsilon$ = 1.8).

\subsection{Results}

\begin{figure}[ht!]
\centering
\includegraphics[width=9cm]{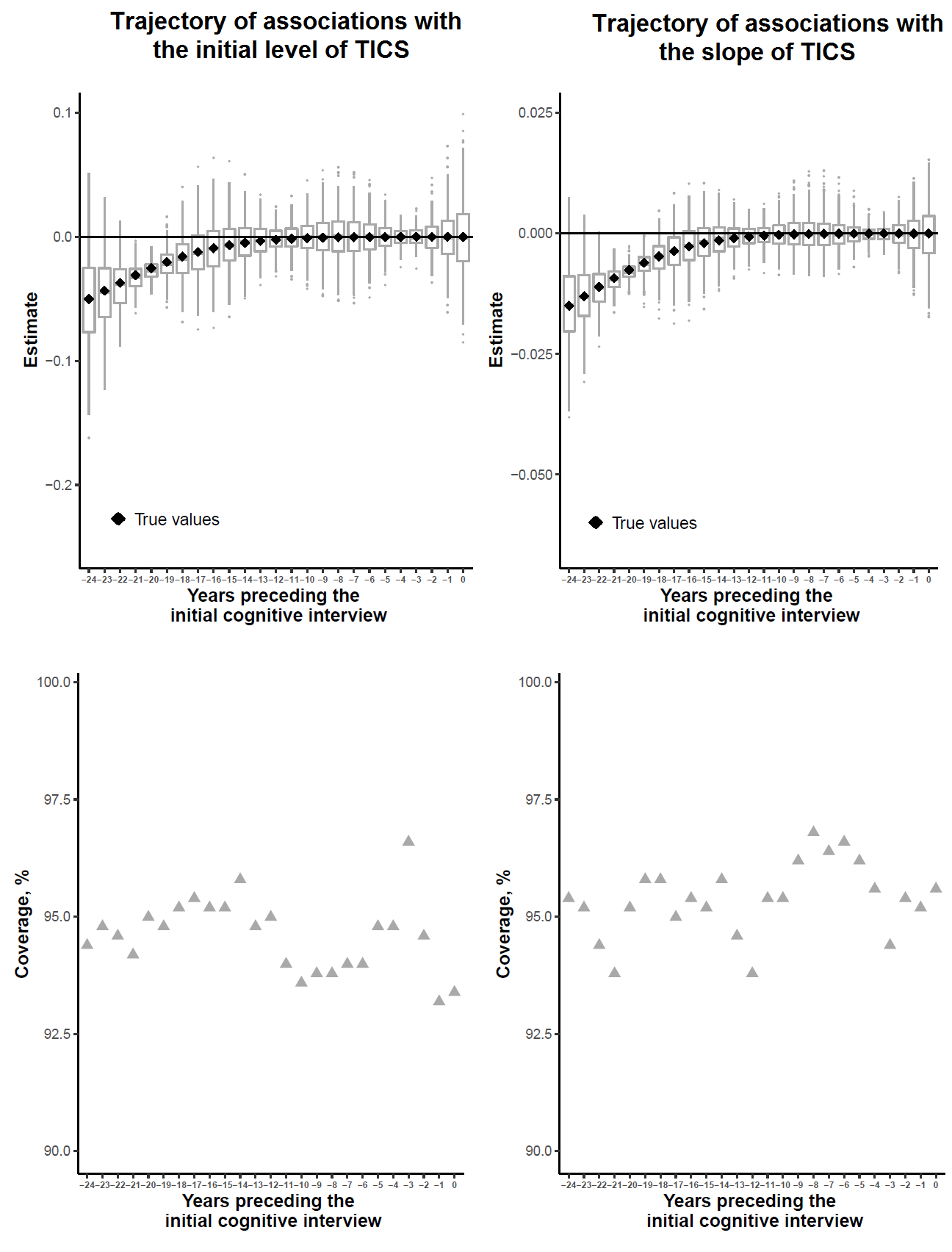}
\caption{\small Boxplots of the trajectory of association between the exposure history over the 24 years prior to the initial health outcome assessment on the initial level (top left panel) or the slope (top right panel) of the outcome of interest across 500 simulations of 1,000 subjects each, and corresponding coverage rates of the 95\% pointwise confidence interval (lower panels) for Scenario B (distant negative effect).}
\label{Figure6}
\end{figure}

We systematically considered 500 replicates of samples of 1,000 participants each. For each scenario, we focused on the estimated trajectories of association with both the initial level and the slope of the repeated outcome, and on the coverage rate of its pointwise 95\% confidence interval obtained by parametric boostrap. As shown in Figure \ref{Figure6} for Scenario B, the two-stage estimation approach retrieved without bias the true generated trajectory of association, and the parametric Boostrap provided satisfying coverage rate of the 95\% confidence interval around the 95\% nominal value. The same conclusions were drawn for Scenario A and C with other shapes of trajectories of association (see eFigure 2 and eFigure 3, respectively).

Furthermore, when considering less repeated measurement points for exposure (see eFigures 4, 5, 6), higher rate of missing data (see eFigures 7, 8, 9), higher measurement error (see eFigures 10, 11, 12), or higher number of inner knots of cubic splines in the definition of the BMI history (see eFigure 13), parameters were again well estimated with negligible bias and no departure from the expected 95\% coverage rate of the 95\% conﬁdence interval. Overall the simulation study demonstrated that in this specific temporal landmark framework, the two-stage procedure combined with parametric bootstrap provided a correct inference.

\newpage
\section*{Discussion}
We proposed a flexible approach for estimating the trajectory of association between a time-varying exposure and a subsequent health outcome when exposure and outcome are assessed non-concurrently. We relied for that on the WCIE methodology (Breslow et al., 1983, Thomas et al., 1988) that we incorporated within a landmark approach to handle the non-concurrent windows for the exposure and outcomes. We followed previous works in WCIE and directly estimated the weights assigned to past exposures from the data using flexible natural cubic splines (Wood et al., 2017). However, in contrast with most previous WCIE approaches that are restricted to complete error-free exposure, our approach considers that the exposure is endogenous, prone to measurement error, and assessed at discrete and individual-specific times; this naturally handles the intermittent missing visits usually encountered in cohort studies and limits the biases induced by neglecting the measurement error that is most likely present in many epidemiological contexts. Mauff et al. (Mauff et al., 2017) had previously extended WCIE to handle endogenous exposure data. However, their method was limited to concomitant exposure and outcome assessments while in life-long epidemiology, exposures remote from the outcome risk-period are also of great interest. In addition, they had only considered time-to-event outcomes while we wanted to also consider repeated outcomes that are frequently encountered in longitudinal studies. Danieli et al. (Danieli et al., 2020) have just proposed a WCIE methodology for repeated outcome data. However, their method is limited to exogenous complete error-free exposures, and evaluates the cumulative effect of the exposure on the current level of the outcome while in etiological studies such as ours, the interest is usually in the effect of the exposure on the rate of change (Voelkle et al., 2018).

Thanks to the separation between the assessment periods of the exposure and the outcome in the landmark approach, we were able to correctly estimate the time-varying effects of the exposure history using a two-stage approach, as demonstrated by simulations. This two-stage procedure makes our methodology easy to implement in standard statistical software, and easy to adapt to other contexts as long as exposure and outcome are not assessed simultaneously. First, any type of outcome could be considered such as time-to-event or binary endpoint as usually encountered in case-control studies. Note however that (i) in case-control studies (unlike prospective studies or case-control studies nested within a cohort), exposure data are collected retrospectively from the landmark time, when the groups were selected, which may result in non-differential measurement errors; (ii) with a time-to-event endpoint, our approach would still require that the exposure time-window ends at the landmark time.
Second, non-ignorable dropout for the repeated outcome could be easily taken into account by estimating the outcome parameters in the second stage using a shared random-effect model (Rizopoulos et al., 2012) rather than a standard mixed model. Third, with a repeatedly-measured outcome, any relevant nonlinear shape of trajectory over time could be considered instead of a linear trajectory. Fourth, the approach could be easily adapted to handle other types of exposures, such as categorical variables, by considering a generalized linear mixed model in the first stage. Finally, for ease of epidemiological interpretation we defined the WCIE as the sum of weighted exposures at each time unit over the $[-S,0]$ period but the integral over the continuous time window could also be considered. 

Simulation studies showed that the proposed two-stage estimation model recovers various sets of shapes of the true weight function (i.e., constant over time or time-varying), and provides satisfactory estimates of the strength of the associations. In particular, splitting the estimation procedure into two parts does not seem to lead to bias when the exposure and outcome are non-concurrent, even in the case of poorer exposure information (i.e., higher missing data or higher time-interval between two visits or higher measurement error).

The method presents however several limitations that warrant consideration. First, while our estimates are capable of capturing the shape of the weight function, the splines functions may not accurately reflect rapid and sudden changes in the relationship (Sylvestre et al., 2009). If these effects are expected, the user can improve the model by including one or more knot(s) in regions where a strong curvature is anticipated or by considering a piecewise constant trajectory of association instead of splines. Second, our methodology assumes a linear association between the exposure at each time and the outcome. Extensions to account for nonlinear associations both over time and over the exposure range is a direction for future research. Third, as in previous WCIE methods, we only considered one time-varying exposure while it could be of interest in the future to simultaneously examine several exposure histories that are known to be interrelated (e.g., BMI and physical activity).

Applied to the association between BMI history starting at midlife and cognition at older ages, our methodology confirmed the complex and age-dependent relationship. Indeed, with no a priori assumptions on the shape of relations, and after controlling for BMI history, age and educational level, we found that higher BMI levels at midlife but lower BMI levels at older ages were significantly associated with poorer cognition after the age 70 in women. This bidirectional relationship is consistent with previous studies, including our own work, that showed that obesity at middle age but low BMI late in life were associated with subsequent development of cognitive impairment and dementia (Albanese et al., 2017, Singh-Manoux et al., 2018, Solomon et al., 2014, Wagner et al., 2018, Wagner et al., 2019). Note however that our interpretations are based on a specific population of generally healthy women and that these results cannot be generalized to populations with other socioeconomic or health status. Although a non-linear J-shaped association (i.e., underweight and overweight are associated with an increased risk of worst cognition) was sometimes reported for BMI in the literature (Albanese et al., 2017), this is unlikely to be the case in our sample in which very few nurses were underweight during the whole exposure period. Finally, we note that the follow-up rate was high in the NHS and there was little or no attrition so that we did not need to account for death or possible informative dropout in our application. 

\section*{Conclusions}

\noindent This methodology offers great flexibility in capturing complex dynamic relationships over the life-course, but also allows application to the majority of epidemiological contexts when the shape of the effects is not known and even more importantly when the history of exposure is measured incompletely and with error. Overall, this method may significantly contribute to broadening the applications of WCIE for a variety epidemiological contexts.

\section*{Abbreviations}
\noindent BMI, body mass index
\\CIE, cumulative index of exposure
\\NHS, Nurses' Health Study
\\WCIE, weighted cumulative index of exposure

\section*{Acknowledgements}
\noindent We would like to thank the participants and staff of the Nurses’ Health Study for their valuable contributions. The Nurses' Health Study was supported by the National Institutes of Health; grant UM1 CA186107. The study protocol was approved by the institutional review board of the Brigham and Women’s Hospital. This work was funded by the France Alzheimer Association (grant number 1875 for project grant DATALInk) and the French National Research Agency (grant number ANR-18-CE36-0004-01 for project DyMES). Computer time was provided by the computing facilities MCIA (Mésocentre de Calcul Intensif Aquitain) at the University of Bordeaux and the University of Pau and Pays de l’Adour.

\section*{Ethics approval and consent to participate}
The institutional review boards of Brigham and Women’s Hospital and Harvard T.H. Chan School of Public Health approved the study protocol of the Nurses' Health Study. Informed consent was implied through the return of the baseline questionnaire of participating women.

\section*{Authors' contributions}
\noindent MW, KL, CS, and CPL conceived and designed the study; collected the data; analyzed the data; MW, CS, and CPL wrote the paper. MW and FG had full access to all of the data in the study. All authors take responsibility for the integrity of the data and the accuracy of the data analysis. All authors read and approved the final manuscript. 

\section*{Consent for publication}
\noindent All the authors gave their final approval of the version to be published.

\section*{Competing interests}
\noindent The authors declare that they have no competing interests. 

\section*{Funding}
\noindent This work was funded by the French National Research Agency (grant number ANR-18-CE36-0004-01 for project DyMES) and France Alzheimer Association (grant number 1875 for project grant DATALInk).

\section*{Availability of data and materials}
\noindent The data that support the findings of this study are available from the corresponding author upon reasonable request with exception to the application raw data from the Nurses' Health Study. Access to Nurses’ Health Study data can be requested at www.nurseshealthstudy.org. The programs of simulation studies are openly available in GitHub at https://github.com/MaudeWagner/WHistory.

\vspace{6mm}
\section*{REFERENCES}
\vspace{6mm}

Abrahamowicz, M., Bartlett, G., Tamblyn, R., \& du Berger, R. (2006). Modeling cumulative dose and exposure duration provided insights regarding the associations between benzodiazepines and injuries. J Clin Epidemiol, 59(4), 393–403.

Abrahamowicz, M., MacKenzie, T., \& Esdaile, J. (1996). Time-dependent hazard ratio: Modeling and hypothesis testing with application in lupus nephritis. Journal of the Acoustical Society of America, 91, 1432–1439.

Akaike, H. (1978). A bayesian analysis of the minimum a.i.c.. procedure. Ann. Inst. Statist. Math., 30, 9–14.

Albanese, E., Launer, L., Egger, M., Prince, M., Giannakopoulos, P., Wolters, F., \& K, E.(2017). Body mass index in midlife and dementia: Systematic review and meta-regression analysis of 589,649 men and women followed in longitudinal studies .Alzheimers Dement, 8, 165–178.

Boos, D., \& Stefanski, L. (2013). Essential statistical inference: Theory and methods. Springer.

Breslow, N., Lubin, J., Marek, P., \& Langholz, B. (1983). Multiplicative models and cohort analysis. J Am Stat Assoc, 78(381), 1–12.

Checkoway, H., \& Rice, C. (1992). Time-weighted averages, peaks, and other indices ofexposure in occupational epidemiology. Am J Ind Med,21(1), 25–33.

Colditz, G., Manson, J., \& Hankinson, S. (1997). The nurses’ health study: 20-year contribution to the understanding of health among women. J Womens Health, 6(1), 49–62.

Cox, D. (1972). Regression models and life-tables.Journal of the Royal Statistical Sociaty. Series B (Methodological), 34, 187–220.

Efron, B., \& Tibshirani, R. (1993). An introduction to the bootstrap. AAPS J, Boca Raton, Florida: Chapman \& Hall/CRC.

Eilers, H., Marx, B., \& Durb, M. (2015). Twenty years of p-splines. SORT,39(2), 149–186.

Folstein, M., Folstein, S., \& McHugh, P. (1975). mini-mental state ». a practical method for grading the cognitive state of patients for the clinician. J Psychiatr Res, 12(3), 189–198.

Hauptmann, M., Wellmann, J., Lubin, J., Rosenberg, P., \& Kreienbrock, L. (2000). Analysis of exposure-time-response relationships using a spline weight function. Biometrics, 56(4), 1105–1108.

Lacourt, A., Lévêque, E., Guichard, E., Gilg Soit Ilg, A., MP, S., \& Leffondré, K. (2017). Dose-time-response association between occupational asbestos exposure and pleural mesothelioma. Occup Environ Med, 74(9), 691–697.

Laird, N., \& Ware, J. (1982). Random-effects models for longitudinal data. Biometrics, 963–974.

Langholz, B., Thomas, D., Xiang, A., \& Stram, D. (1999). Latency analysis in epidemiologic studies of occupational exposures: Application to the colorado plateau uraniumminers cohort. Am J Ind Med, 35(3), 246–256.

Lévêque, E., Lacourt, A., Luce, D., Sylvestre, M., Guénel, P., Stücker, I., \& Leffondré, K.(2018). Time-dependent effect of intensity of smoking and of occupational exposure to asbestos on the risk of lung cancer: Results from the icare case-control study. Occup Environ Med, 75(8), 586–592.

Liu, S.,  Jones, RN. \& Glymour, MM. (2010). Implications of Lifecourse Epidemiology for Research on Determinants of Adult Disease. Public Health Reviews, 32(2), 489-511.

Mauff, K., Steyerberg, E., Nijpels, G., van der Heijden, A., \& Rizopoulos, D. (2017). Extension of the association structure in joint models to include weighted cumulative effects. Stat Med, 36(23), 3746–3759.

Perperoglou, A., Sauerbrei, W., Abrahamowicz, M., \& Schmid, M. (2019). A review ofspline function procedures in r. BMC Med Res Methodol, 19(1), 46.

Prentice, R. (1982). Covariate measurement errors and parameter estimation in a failure time regression model. Biometrika, 69(2), 331–342.

Proust-Lima, C., Philipps, V., \& Liquet, B. (2017). Estimation of extended mixed modelsusing latent classes and latent processes: The r package lcmm.J Stat Softw Artic,78, 1–56.

Royston, P., Ambler, G., \& Sauerbrei, W. (1999). The use of fractional polynomials to model continuous risk variables in epidemiology. Int J Epidemiol), 28(5), 964–974.

Singh-Manoux, A., Czernichow, S., Elbaz, A., Dugravot, A., Sabia, S., Hagger-Johnson, G.,Kaffashian, S., Zins, M., Brunner, E., Nabi, H., \& Kivimäki, M. (2012). Obesity phenotypes in midlife and cognition in early old age: The whitehall ii cohort study.Neurology,79(8), 755–762.

Singh-Manoux, A., Dugravot, A., Shipley, M., Brunner, E., Elbaz, A., Sabia, S., \& Kivimaki,M. (2018). Obesity trajectories and risk of dementia: 28 years of follow-up in the whitehall II study. Alzheimers Dement,14(2), 178–186.

Smith, P. (1979). Splines as a useful and convenient statistical tool. American Statistician,33(2), 57–62.

Solomon, A., Mangialasche, F., Richard, E., Andrieu, S., Bennett, D., Breteler, M., Fratiglioni,L., Hooshmand, B., Khachaturian, A., Schneider, L., Skoog, I., \& M, K. (2014). Advances in the prevention of alzheimer’s disease and dementia. J. Intern. Med.,275,229–250.

Stranges, S., Bonner, M., Fucci, F., Cummings, K., Freudenheim, J., Dorn, J., \& et al. (2006). Lifetime cumulative exposure to secondhand smoke and risk of myocardial infarction in never smokers: Results from the western new york health study, 1995-2001.Arch Intern Med,166(18), 1961–1967.

Sylvestre, M., \& Abrahamowicz, M. (2009). Flexible modeling of the cumulative effects of time-dependent exposures on the hazard. Stat Med,28(27), 3437–3453.

Sylvestre, M., Abrahamowicz, M., Čape, R., \& Tamblyn, R. (2012). Assessing the cumulative effects of exposure to selected benzodiazepines on the risk of fall-related injuries in the elderly. Int Psychogeriatr,24(4), 577–586.

Thomas, D. (1988). Models for exposure-time-response relationships with applications to cancer epidemiology. Annu Rev Public Health, (9), 451–482.

Tsiatis, A., \& Davidian, M. (2004). Joint modeling of longitudinal and time-to-event data: An overview. Statistica Sinica,14(3), 809–834.

Vaceck, P. (1997). Assessing the effect of intensity when exposure varies over time. StatMed,16(5), 505–513.

Vivot, A., Power, M., \& Glymour, M. (2016). Jump, hop, or skip: Modeling practice effectsin studies of determinants of cognitive change in older adults. Am J Epidemiol,183(4), 302–314.

Wagner, M., Grodstein, F., Proust-Lima, C., \& Samieri, C. (2020). Long-term trajectories of body weight, diet, and physical activity from midlife through late-life and subsequent cognitive decline in women. American Journal of Epidemioly, 189(4):305-313.

Wagner, M., Helmer, C., Tzourio, C., Berr, C., Proust-Lima, C., \& Samieri, C. (2018). Evaluation of the concurrent trajectories of cardiometabolic risk factors in the 14 years before dementia. JAMA Psychiatry, 75(10), 1033–1042.

Wang, M., Liao, X., Laden, F., \& Spiegelman, D. (2016). Quantifying risk over the life course- latency, age-related susceptibility, and other time-varying exposure metrics. StatMed,35(13), 2283–2295.

West, N., Lirette, S., Cannon, V., Turner, S., Mosley, T. J., \& Windham, B. (2017). Adiposity, change in adiposity, and cognitive decline in mid- and late life. J Am GeriatrSoc,65(6), 1282–1288.

Willett, W., Stampfer, M., Bain, C., Lipnick, R., Speizer, F., Rosner, B., \& et al. (1983). Cigarette smoking, relative weight, and menopause. Am J Epidemiol, 117(6), 651–658.

Wood, S. (2017).Generalized additive models: An introduction with r. Chapman; Hall/CRC.

\newpage
\section*{SUPPLEMENTARY MATERIALS}

\begin{figure}[ht!]
\centering
\includegraphics[width=11cm]{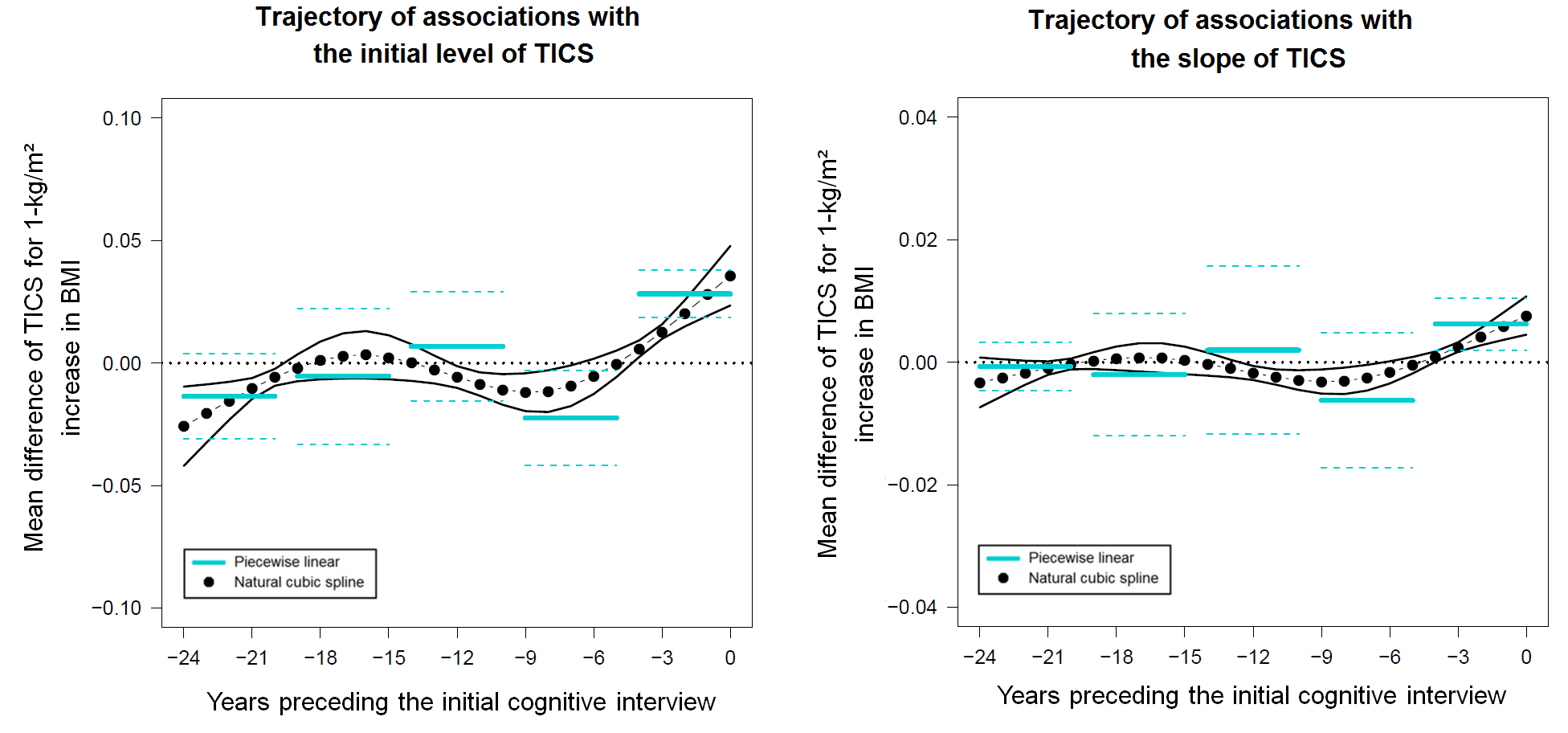}
\legend{\textbf{eFigure 1.} Trajectories of associations between the body mass index history in the 24 years prior to the first cognitive interview on the initial level (left panel) or the slope (right panel) of the Telephone Interview for Cognitive Status (TICS) score approximated by natural cubic splines (in black) or 5-year piecewise constants (in blue) in the Nurses' Health Study (N=19,381), United States (1976-2000). 95\% confidence intervals were obtained by parametric bootstrap with 500 replicates}
\label{eFigure1}
\end{figure}

\begin{figure}[ht]
\centering
\includegraphics[width=12cm]{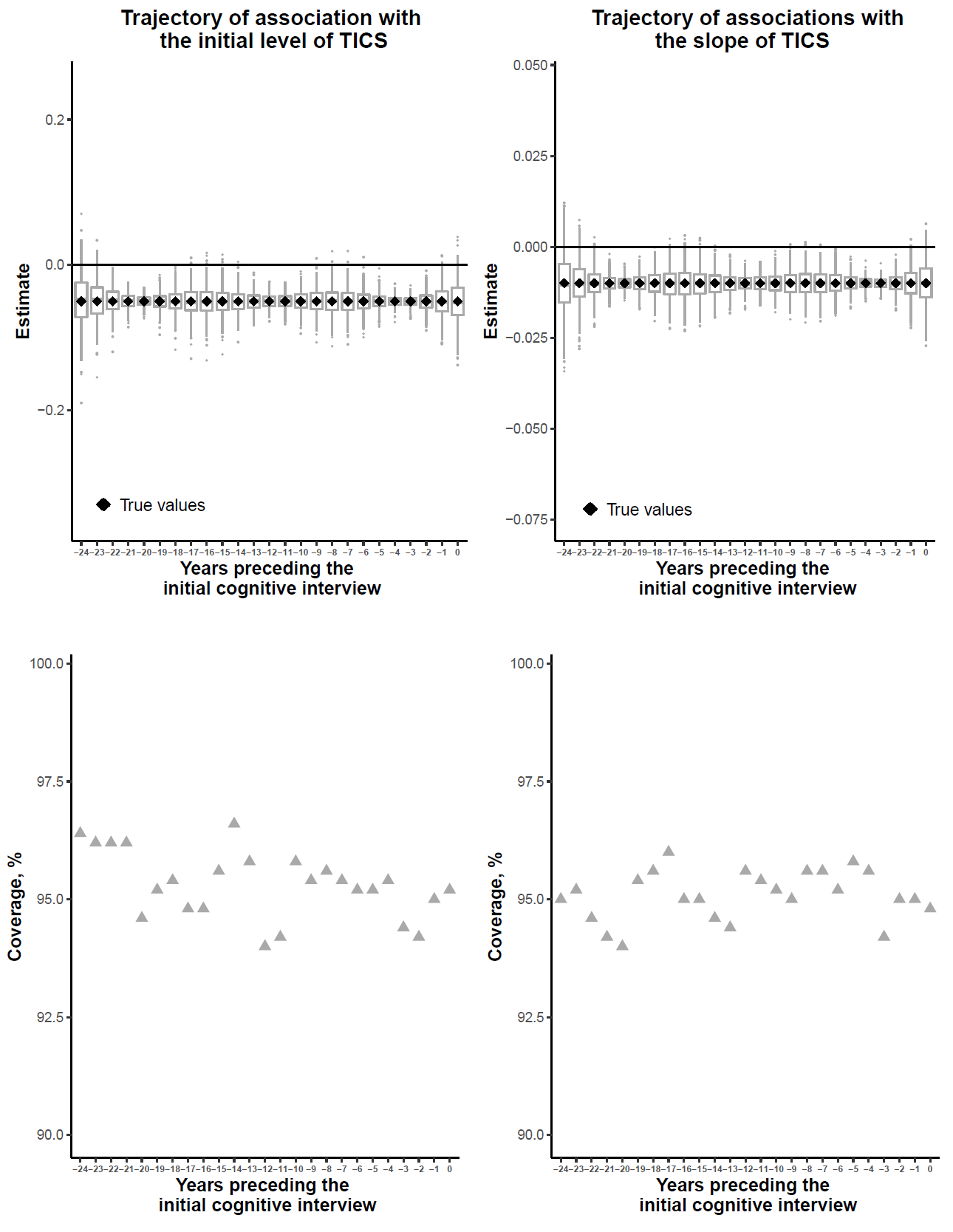}
\legend{\textbf{eFigure 2.} Boxplots of the trajectory of association between the exposure history over the 24 years prior to the initial health outcome assessment on the initial level (top left panel) and slope (top right panel) of the outcome of interest across 500 simulations of 1,000 subjects each, and corresponding coverage rates (low panels) for Scenario A (constant effect).}
\label{eFigure2}
\end{figure}

\begin{figure}[ht]
\centering
\includegraphics[width=12cm]{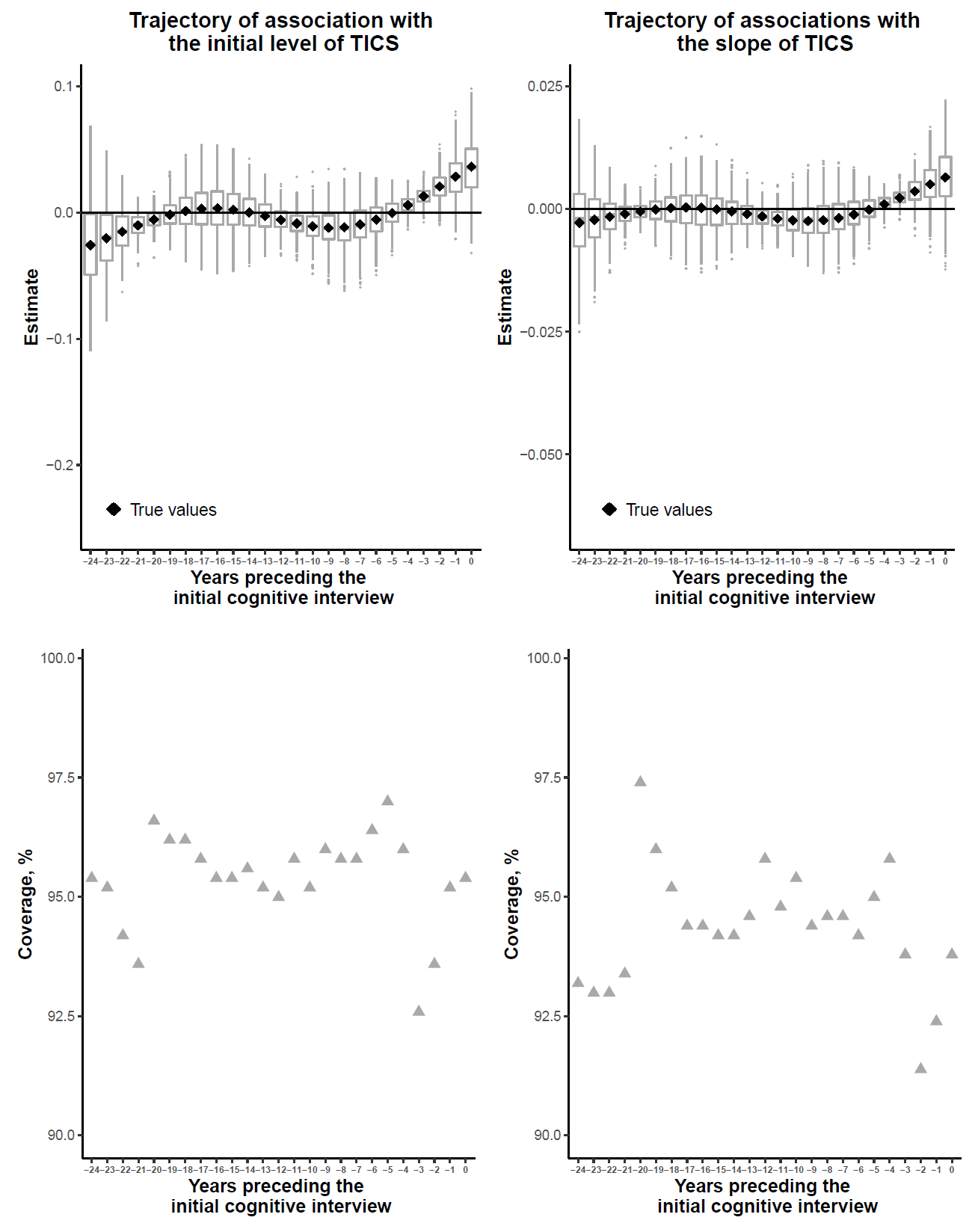}
\legend{\textbf{eFigure 3.} Boxplots of the trajectory of association between the exposure history over the 24 years prior to the initial health outcome assessment on the initial level (top left panel) and slope (top right panel) of the outcome of interest across 500 simulations of 1,000 subjects each, and corresponding coverage rates (low panels) for Scenario C (effect mimicking the associations between BMI and TICS in the Nurses’ Health Study).}
\label{eFigure3}
\end{figure}


\begin{figure}[ht]
\centering
\includegraphics[width=12cm]{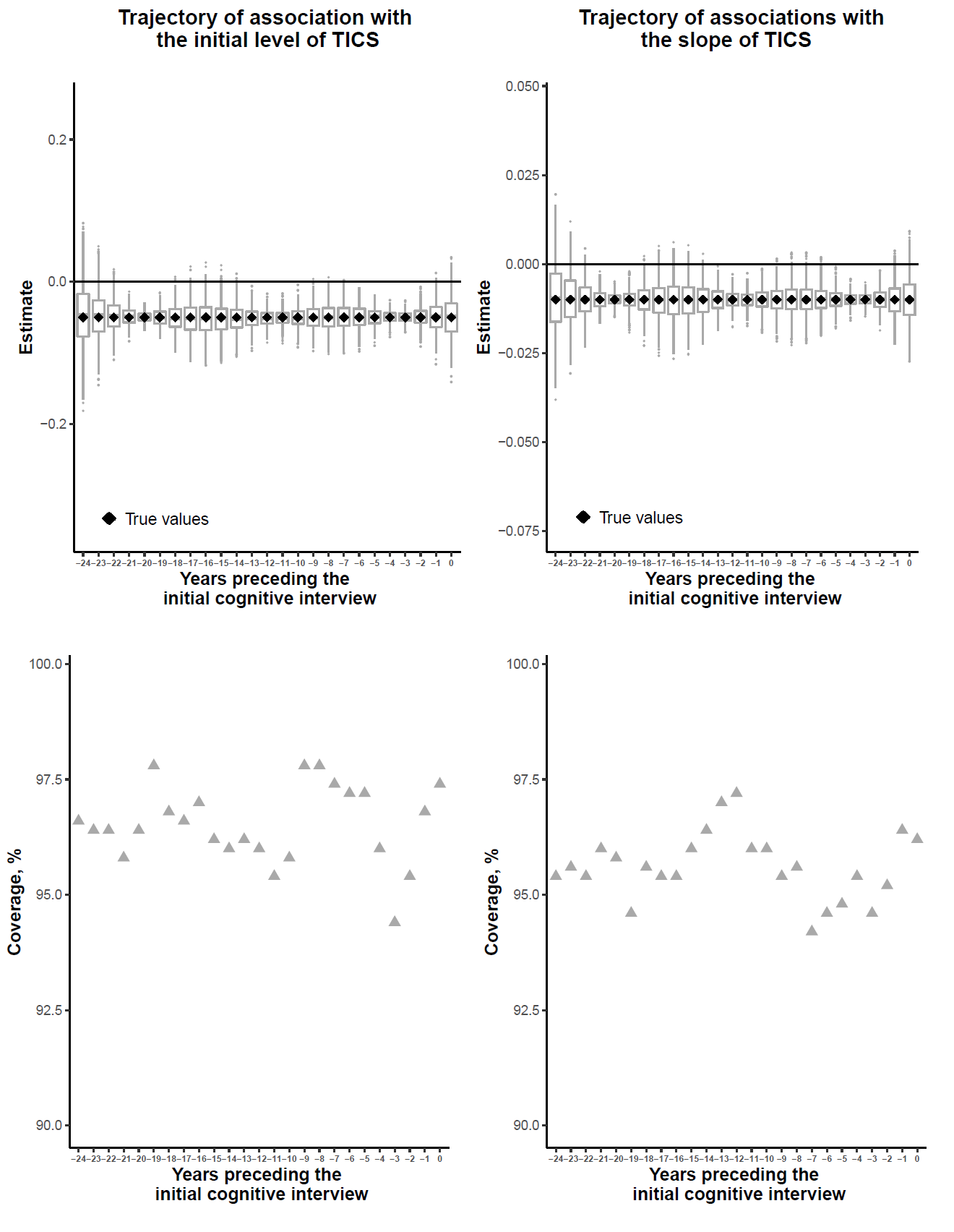}
\legend{\textbf{eFigure 4.} Boxplots of the trajectory of association between the exposure history over the 24 years prior to the initial health outcome assessment on the initial level (top left panel) and slope (top right panel) of the outcome of interest when considering less repeated information for exposure (i.e., measured every 4 years instead of every 2 years) across 500 simulations of 1,000 subjects each, and corresponding coverage rates (low panels) for Scenario A (constant effect).}
\label{eFigure4}
\end{figure}

\begin{figure}[ht]
\centering
\includegraphics[width=12cm]{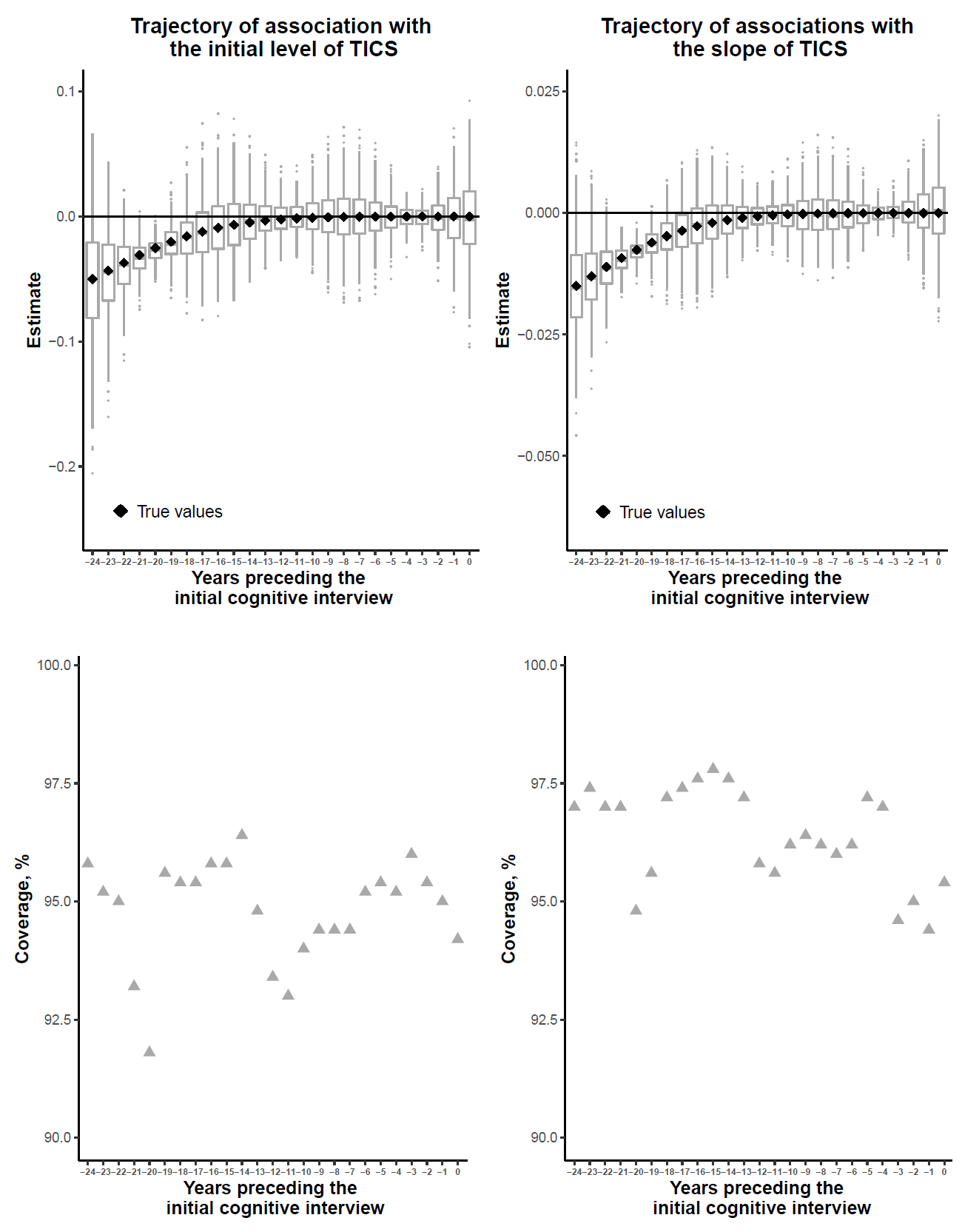}
\legend{\textbf{eFigure 5.} Boxplots of the trajectory of association between the exposure history over the 24 years prior to the initial health outcome assessment on the initial level (top left panel) and slope (top right panel) of the outcome of interest when considering less repeated information for exposure (i.e., measured every 4 years instead of every 2 years) across 500 simulations of 1,000 subjects each, and corresponding coverage rates (low panels) for Scenario B (distant negative effect).}
\label{eFigure5}
\end{figure}

\begin{figure}[ht]
\centering
\includegraphics[width=12cm]{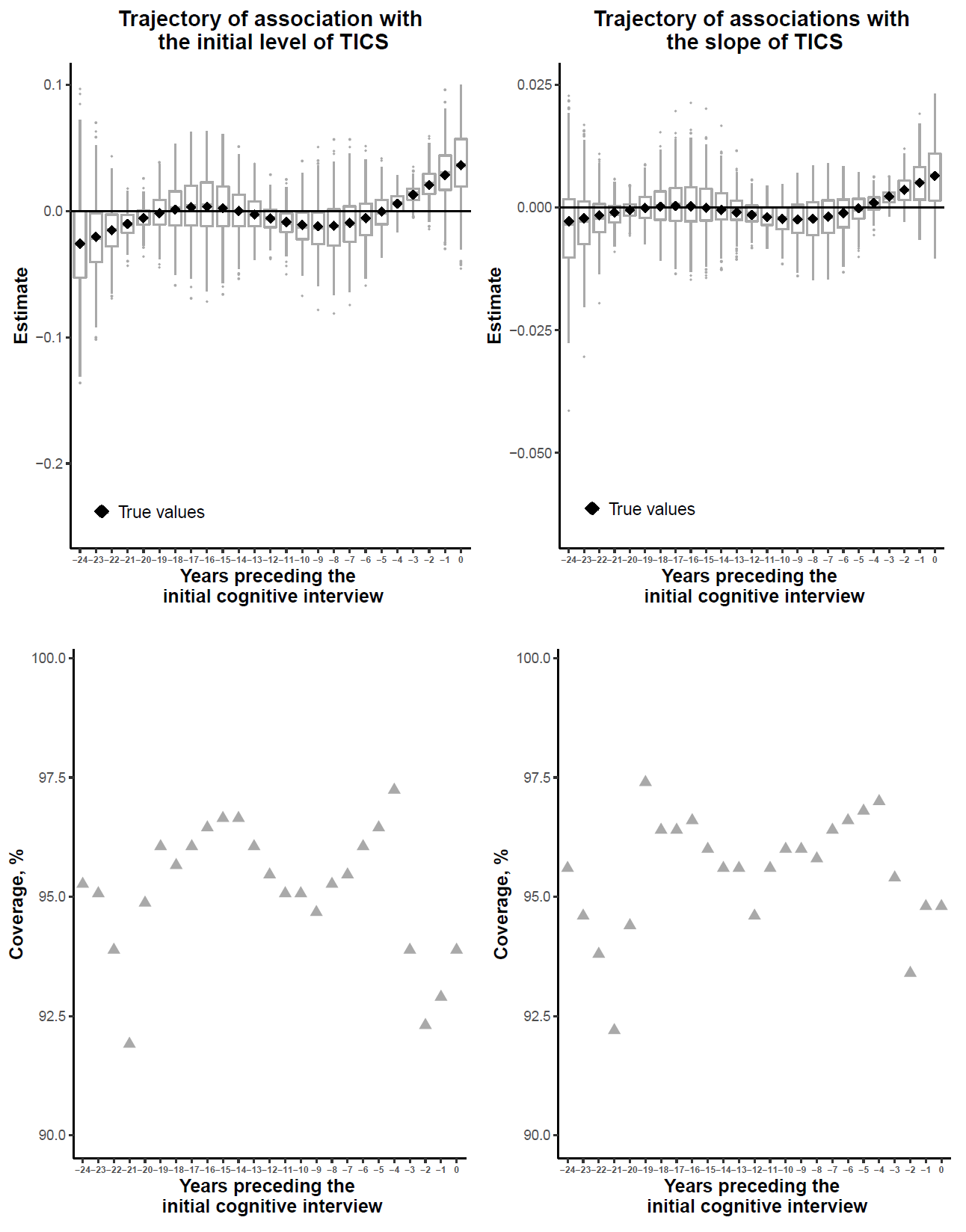}
\legend{\textbf{eFigure 6.} Boxplots of the trajectory of association between the exposure history over the 24 years prior to the initial health outcome assessment on the initial level (top left panel) and slope (top right panel) of the outcome of interest when considering less repeated information for exposure (i.e., measured every 4 years instead of every 2 years) across 500 simulations of 1,000 subjects each, and corresponding coverage rates (low panels) for Scenario C (effect mimicking the associations between BMI and TICS in the Nurses’ Health Study).}
\label{eFigure6}
\end{figure}


\begin{figure}[ht]
\centering
\includegraphics[width=12cm]{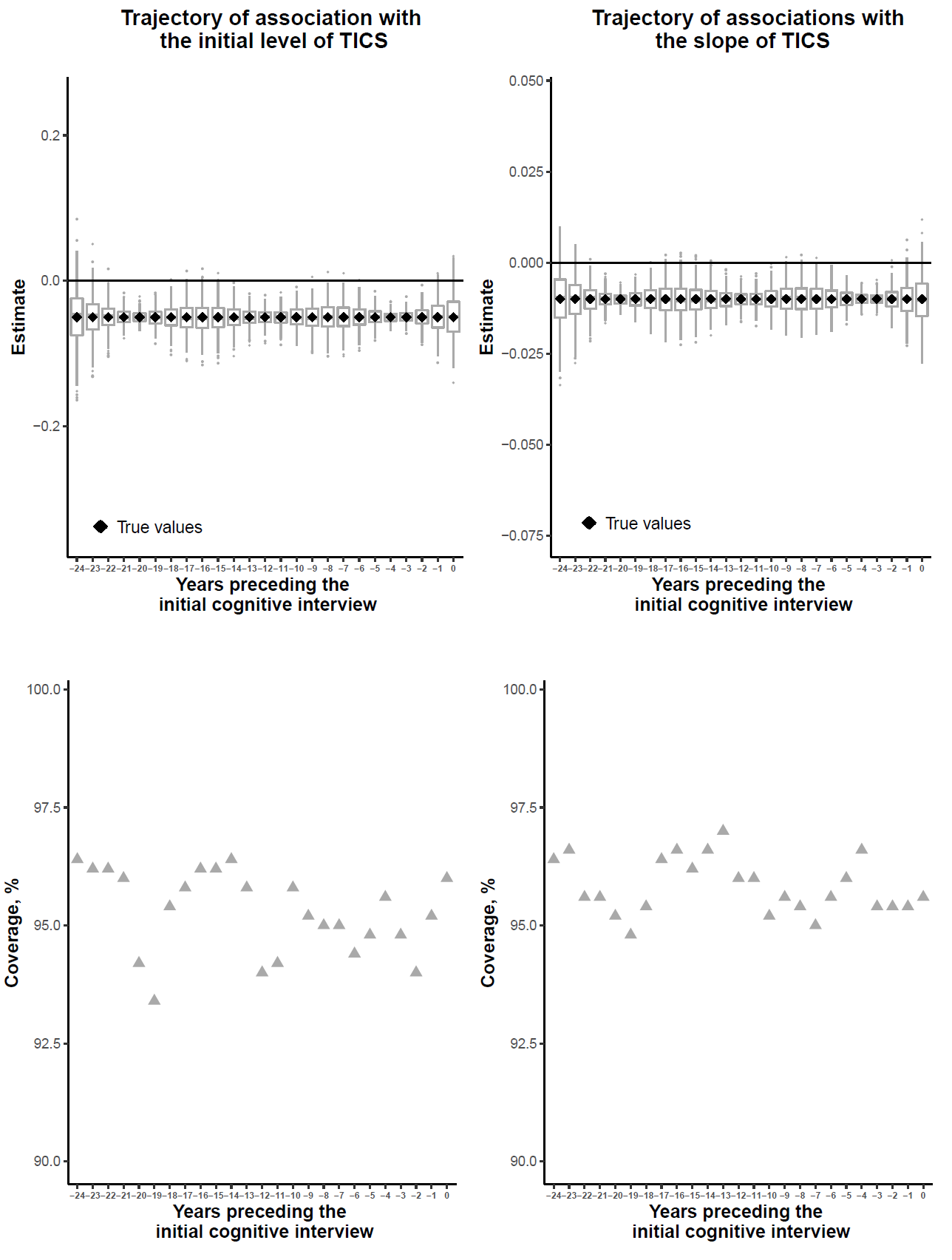}
\legend{\textbf{eFigure 7.} Boxplots of the trajectory of association between the exposure history over the 24 years prior to the initial health outcome assessment on the initial level (top left panel) and slope (top right panel) of the outcome of interest when considering a larger proportion of missing data for the exposure (i.e., 20\% instead of 10\%) across 500 simulations of 1,000 subjects each, and corresponding coverage rates (low panels) for Scenario A (constant effect).}
\label{eFigure7}
\end{figure}

\begin{figure}[ht]
\centering
\includegraphics[width=12cm]{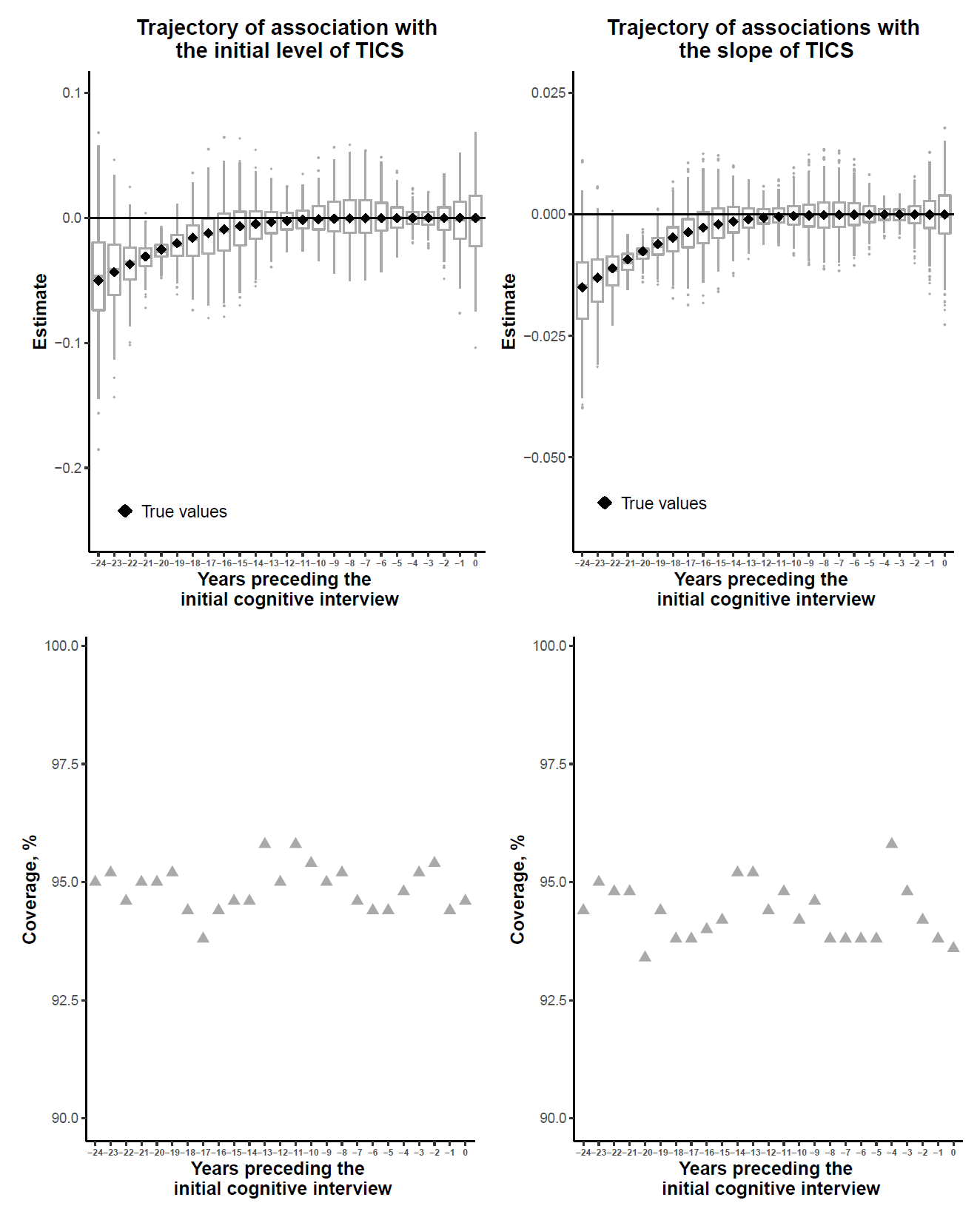}
\legend*{\textbf{eFigure 8.} Boxplots of the trajectory of association between the exposure history over the 24 years prior to the initial health outcome assessment on the initial level (top left panel) and slope (top right panel) of the outcome of interest when considering a larger proportion of missing data for the exposure (i.e., 20\% instead of 10\%) across 500 simulations of 1,000 subjects each, and corresponding coverage rates (low panels) for Scenario B (distant negative effect).}
\label{eFigure8}
\end{figure}

\begin{figure}[ht]
\centering
\includegraphics[width=12cm]{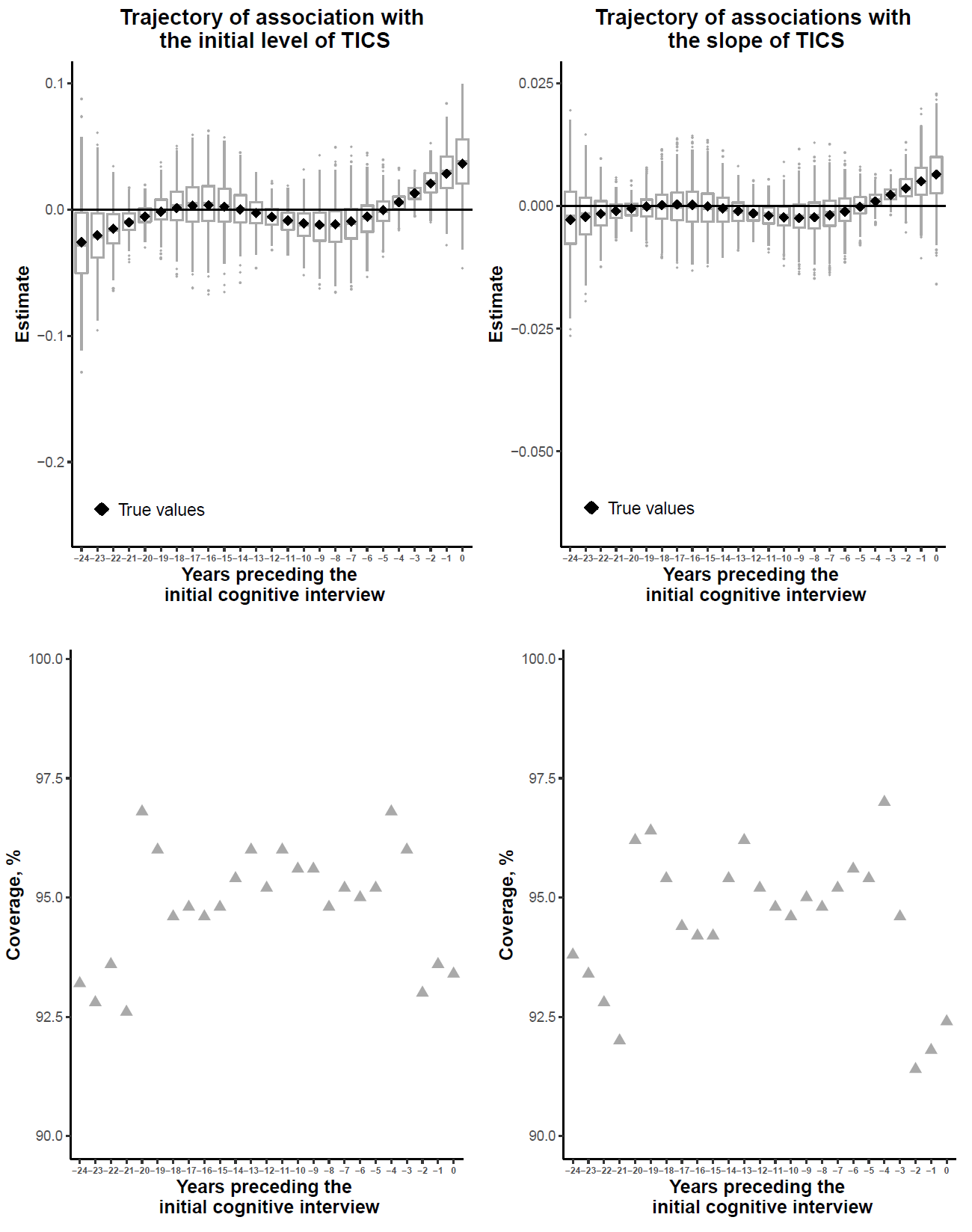}
\legend{\textbf{eFigure 9.} Boxplots of the trajectory of association between the exposure history over the 24 years prior to the initial health outcome assessment on the initial level (top left panel) and slope (top right panel) of the outcome of interest when considering a larger proportion of missing data for the exposure (i.e., 20\% instead of 10\%) across 500 simulations of 1,000 subjects each, and corresponding coverage rates (low panels) for Scenario C (effect mimicking the associations between BMI and TICS in the Nurses’ Health Study).}
\label{eFigure9}
\end{figure}


\begin{figure}[ht]
\centering
\includegraphics[width=12cm]{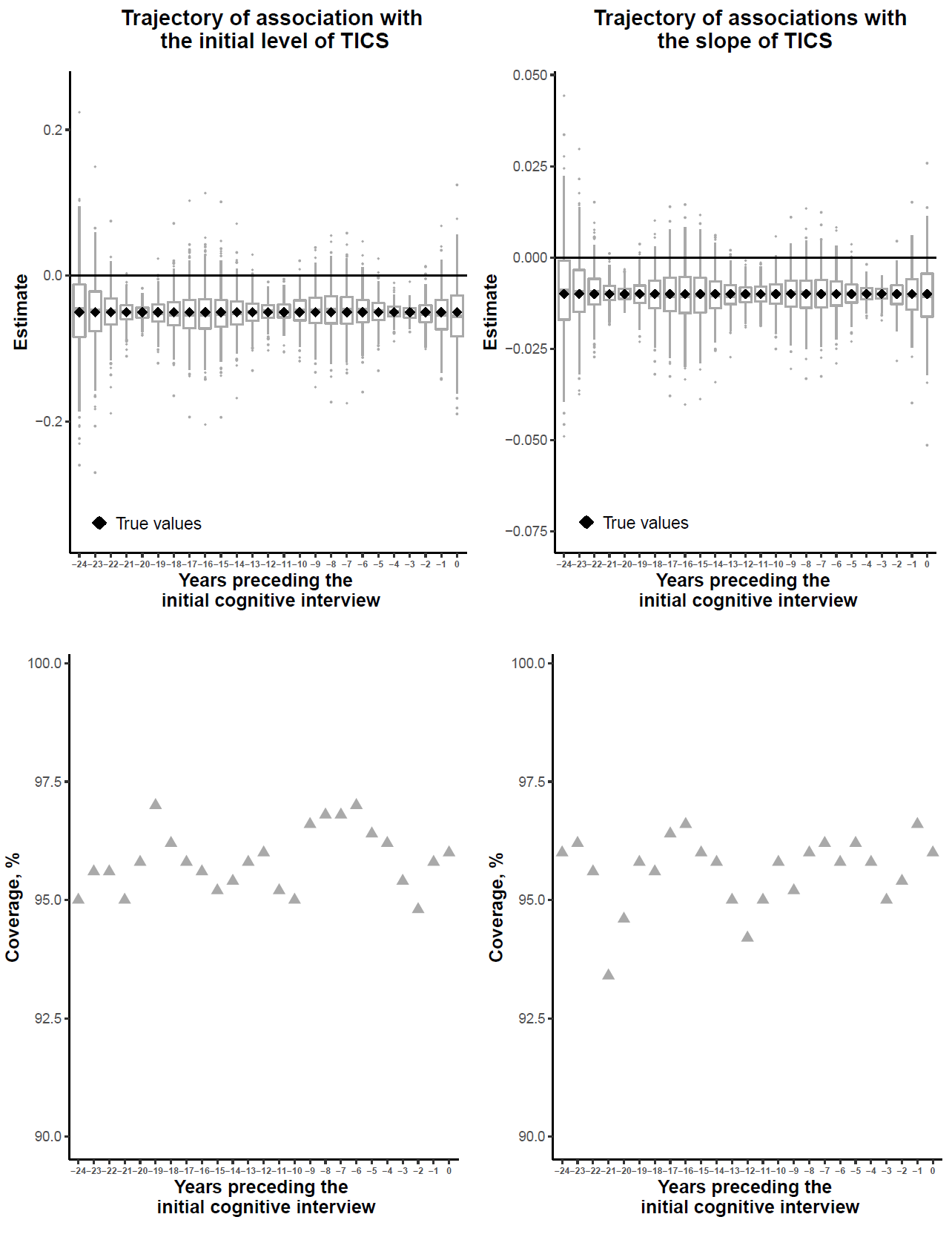}
\legend{\textbf{eFigure 10.} Boxplots of the trajectory of association between the exposure history over the 24 years prior to the initial health outcome assessment on the initial level (top left panel) and slope (top right panel) of the outcome of interest when considering a larger error of measurement (i.e., $\sigma_E$ = 1.8 instead of 0.9) across 500 simulations of 1,000 subjects each, and corresponding coverage rates (low panels) for Scenario A (constant effect).}
\label{eFigure10}
\end{figure}

\begin{figure}[ht]
\centering
\includegraphics[width=12cm]{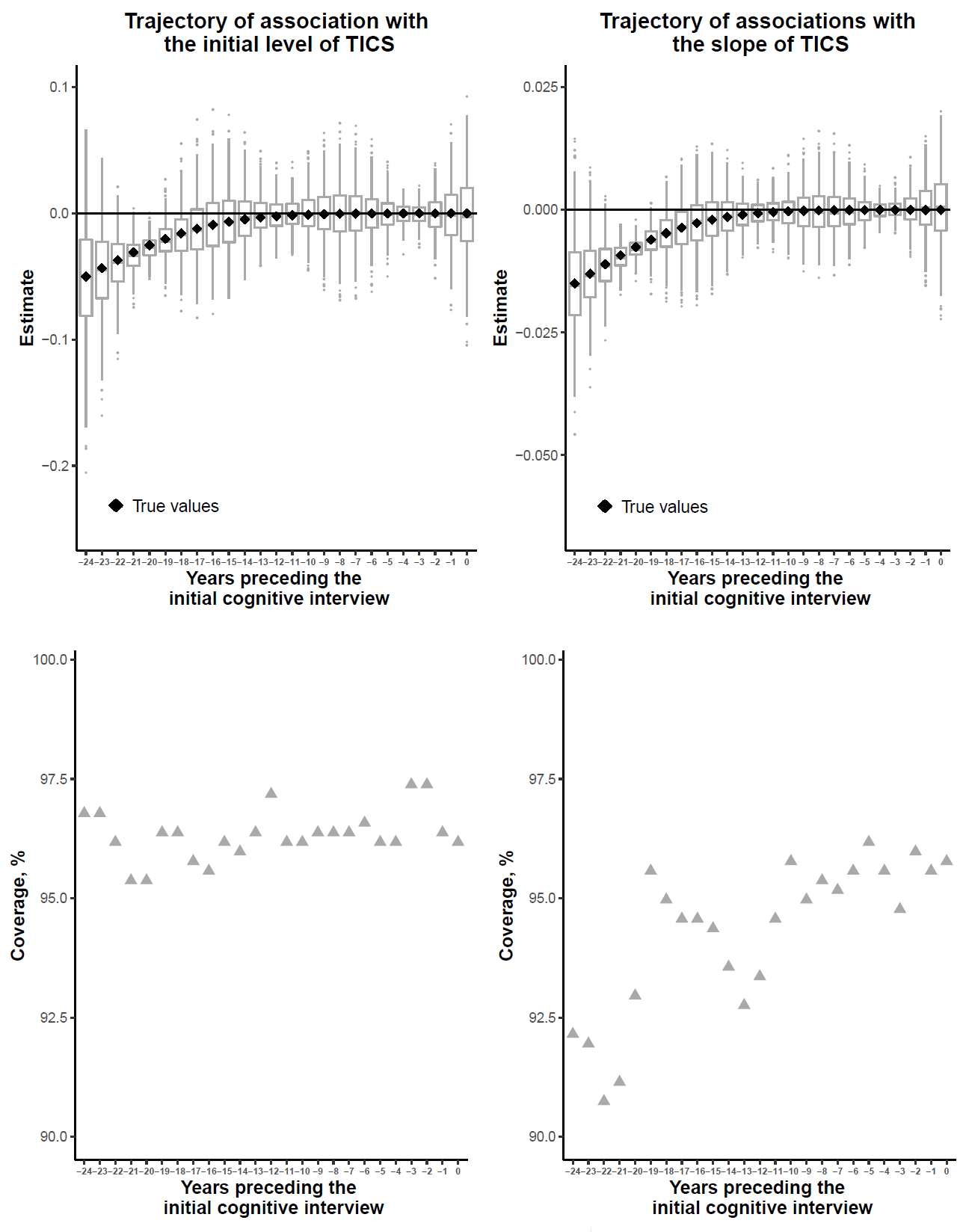}
\legend{\textbf{eFigure 11.} Boxplots of the trajectory of association between the exposure history over the 24 years prior to the initial health outcome assessment on the initial level (top left panel) and slope (top right panel) of the outcome of interest when considering a larger error of measurement (i.e., $\sigma_E$ = 1.8 instead of 0.9) across 500 simulations of 1,000 subjects each, and corresponding coverage rates (low panels) for Scenario B (distant negative effect).}
\label{eFigure11}
\end{figure}

\begin{figure}[ht]
\centering
\includegraphics[width=12cm]{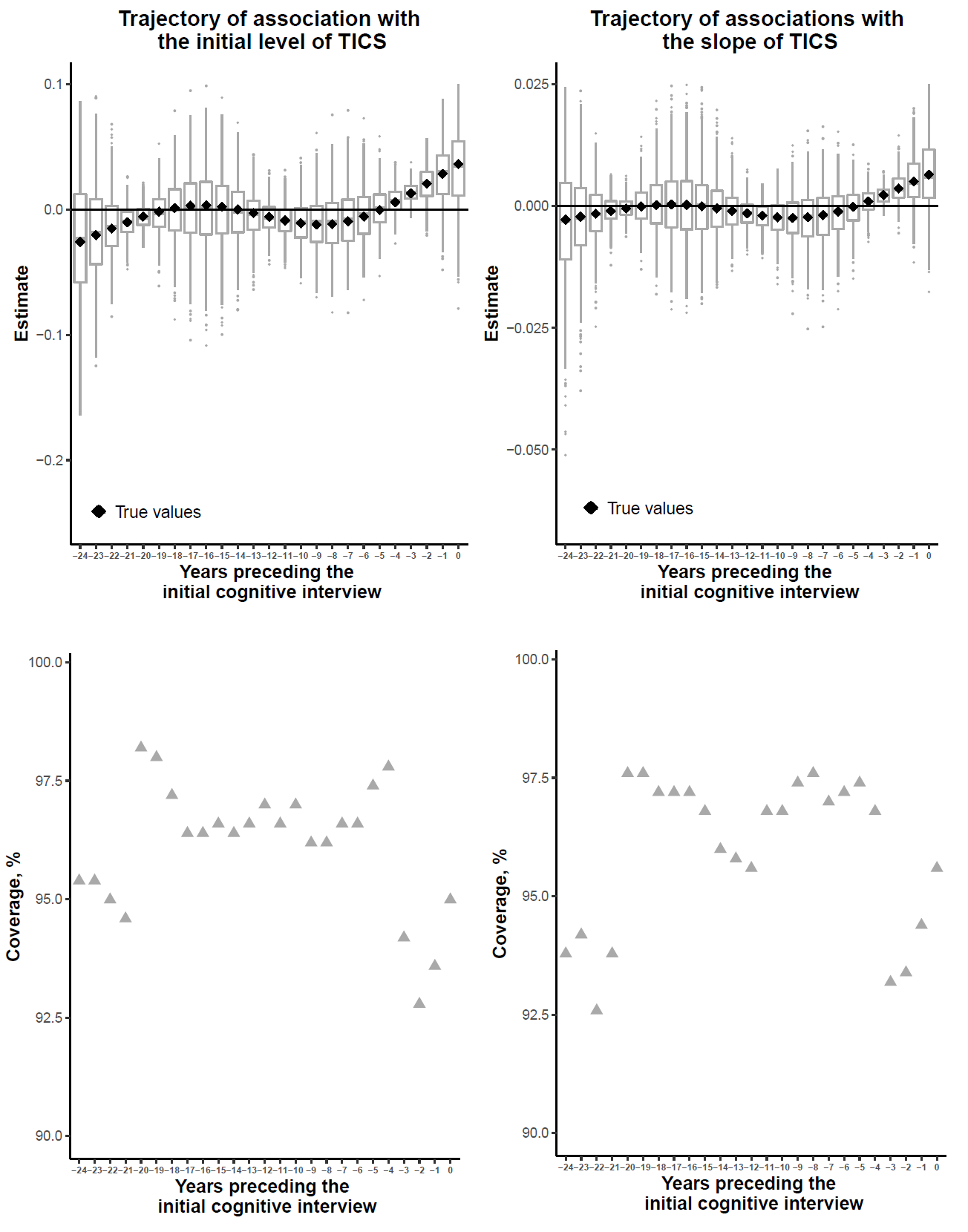}
\legend{\textbf{eFigure 12.} Boxplots of the trajectory of association between the exposure history over the 24 years prior to the initial health outcome assessment on the initial level (top left panel) and slope (top right panel) of the outcome of interest when considering a larger error of measurement (i.e., $\sigma_E$ = 1.8 instead of 0.9) across 500 simulations of 1,000 subjects each, and corresponding coverage rates (low panels) for Scenario C (effect mimicking the associations between BMI and TICS in the Nurses’ Health Study).}
\label{eFigure12}
\end{figure}

\begin{figure}[ht]
\centering
\includegraphics[width=12cm]{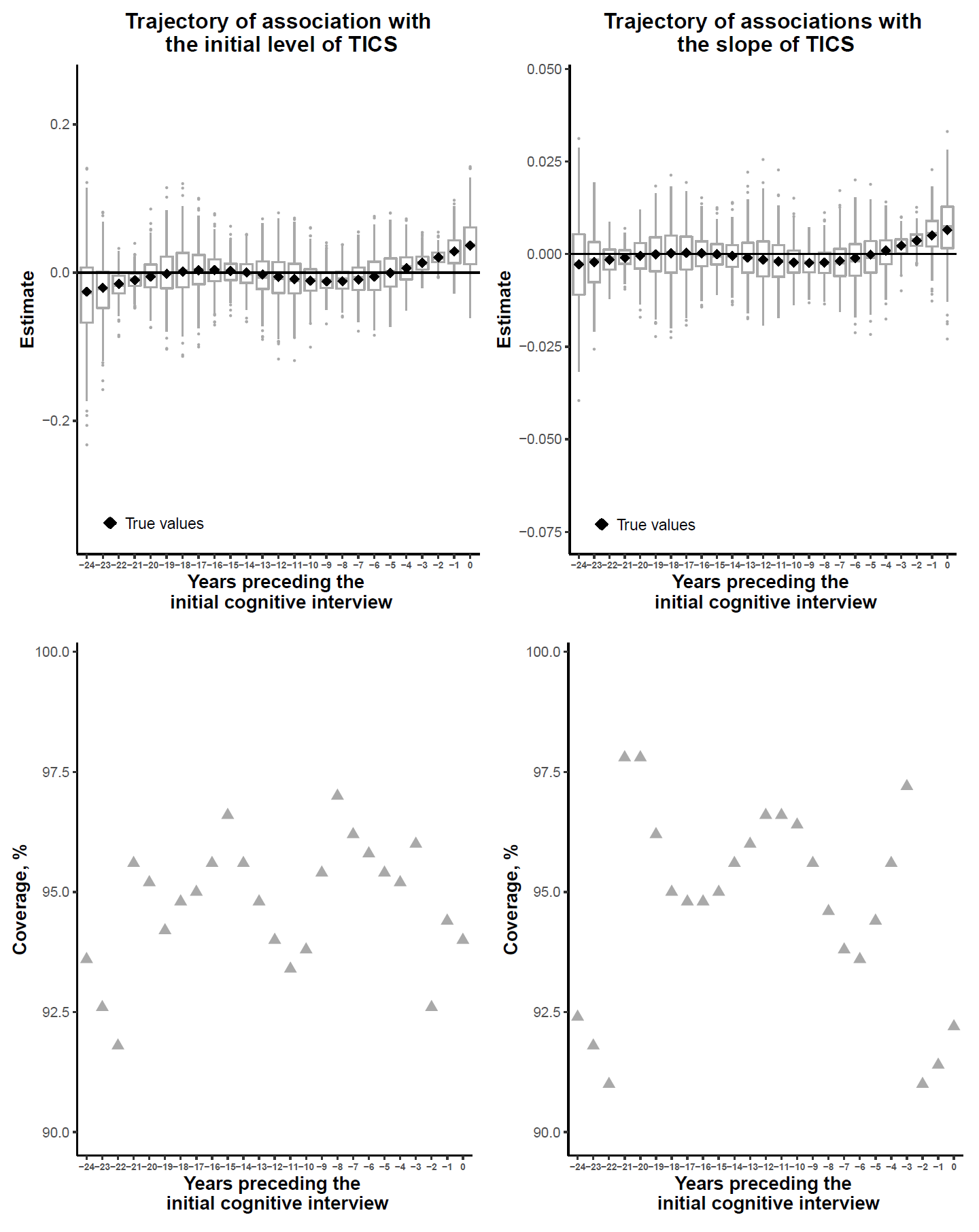}
\legend{\textbf{eFigure 13.} Boxplots of the trajectory of association between the exposure history over the 24 years prior to the initial health outcome assessment on the initial level (top left panel) and slope (top right panel) of the outcome of interest when considering a higher number of inner knots of cubic splines in the definition of the BMI history (i.e., 3 inner knots located at the 25th, 50th, and 75th percentiles instead of 2 located at the 33th and 66th percentiles) across 500 simulations of 1,000 subjects each, and corresponding coverage rates (low panels) for Scenario C (effect mimicking the associations between BMI and TICS in the Nurses’ Health Study).}
\label{eFigure13}
\end{figure}

\end{document}